\documentclass[sigconf]{acmart}

\AtBeginDocument{%
  \providecommand\BibTeX{{%
    \normalfont B\kern-0.5em{\scshape i\kern-0.25em b}\kern-0.8em\TeX}}}

\copyrightyear{2021}
\acmYear{2021}
\setcopyright{acmcopyright}\acmConference[SIGMOD '21]{Proceedings of the 2021
International Conference on Management of Data}{June 20--25, 2021}{Virtual Event,
China}
\acmBooktitle{Proceedings of the 2021 International Conference on Management of
Data (SIGMOD '21), June 20--25, 2021, Virtual Event, China}
\acmPrice{15.00}
\acmDOI{10.1145/3448016.3457327}
\acmISBN{978-1-4503-8343-1/21/06}

\usepackage[utf8]{inputenc} 
\usepackage[T1]{fontenc}    
\usepackage{hyperref}       
\usepackage{url}            
\usepackage{booktabs}       
\usepackage{amsfonts}       
\usepackage{nicefrac}       
\usepackage{microtype}      
\usepackage{appendix}
\usepackage[figuresright]{rotating}
\usepackage{amsmath,amscd,amsthm}
\usepackage{graphicx}
\usepackage{dcolumn}
\usepackage{bm}
\usepackage{ifthen}
\usepackage{rays_defs_18}
\usepackage{commath}
\usepackage{bbm}
\usepackage{wrapfig}
\usepackage{algorithm,algorithmic}
\usepackage{xcolor}
\usepackage{mathtools}
\usepackage[font=small,skip=0pt]{caption}
\usepackage{subcaption}
\usepackage{enumitem}
\usepackage{multirow}
\usepackage{multicol}

\newcommand{\ignore}[1]{}

\usepackage{hyperref}

\newtheorem{lemma}{Lemma}

\newtheorem{theorem}{Theorem}

\newcommand\median{\mathrm{median}}

\definecolor{DarkBlue}{rgb}{0,0,0.7} 

\long\def\comment#1{}

\title{Active Sampling Count Sketch (ASCS) for Online Sparse Estimation of a Trillion Scale Covariance Matrix}

\author{Zhenwei Dai}
\email{zd11@rice.edu}
\affiliation{%
  \institution{Rice University}
  \city{Houston}
  \state{Texas}
  \country{USA}
}
\author{Aditya Desai}
\email{apd10@rice.edu}
\affiliation{%
  \institution{Rice University}
  \city{Houston}
  \state{Texas}
  \country{USA}
}
\author{Reinhard Heckel}
\email{reinhard.heckel@gmail.com }
\affiliation{%
  \institution{Technical University of Munich}
  \city{Munich}
  \country{Germany}
}
\author{Anshumali Shrivastava}
\email{anshumali@rice.edu}
\affiliation{%
  \institution{Rice University}
  \city{Houston}
  \state{Texas}
  \country{USA}
}

\begin{document}
\fancyhead{}
\begin{abstract}
    Estimating and storing the covariance (or correlation) matrix of high-dimensional data is computationally challenging because both memory and computational requirements scale quadratically with the dimension. Fortunately, high-dimensional covariance matrices as observed in text, click-through, meta-genomics datasets, etc are often sparse. In this paper, we consider the problem of efficient sparse estimation of covariance matrices with possibly trillions of entries. The size of the datasets we target requires the algorithm to be online, as more than one pass over the data is prohibitive. In this paper, we propose Active Sampling Count Sketch (ASCS), an online and one-pass sketching algorithm, that recovers the large entries of the covariance matrix accurately. Count Sketch (CS), and other sub-linear compressed sensing algorithms, offer a natural solution to the problem in theory. However, vanilla CS does not work well in practice due to a low signal-to-noise ratio (SNR). At the heart of our approach is a novel active sampling strategy that increases the SNR of classical CS. We demonstrate the practicality of our algorithm with synthetic data and real-world high dimensional datasets. ASCS significantly improves over vanilla CS, demonstrating the merit of our active sampling strategy. 
\end{abstract}

\maketitle

\section{Introduction}

Covariance matrix estimation is a key component of multivariate analysis and machine learning algorithms like Principle Component Analysis (PCA)~\cite{zou2006sparse}) and Canonical Correlation Analysis (CCA)~\cite{hardoon2004canonical}, which are widely used to analyze data. Covariance matrix estimation also has applications in many other areas, including in genomics and climate science. For example, gene association networks are  inferred from empirical covariance matrices~\cite{schafer2005empirical}, and the analysis of inter-annual climate changes heavily relies on estimating the covariances~\cite{yettella2018ensemble}.

\textbf{Problem Setting:}
With the advent of big data, large-scale high dimensional datasets are pervasive throughout many applications. For example, the Large Scale Metagenomic Sequence Dataset ~\cite{vervier2016large} has 16 million features and 100 million samples. For this dataset, storing the covariance matrix (hundreds of terabytes) or even loading the whole dataset into RAM (hundreds of gigabytes) is prohibitive for most machines. This challenge calls for an \textit{online algorithm} that optimizes data processing and a \textit{sketch data structure}, which compresses the matrix to preserve the signals of interest, which we consider to be the large entries of the matrix.

\begin{figure}[ht]
\centering 
\includegraphics[width=0.3 \textwidth]{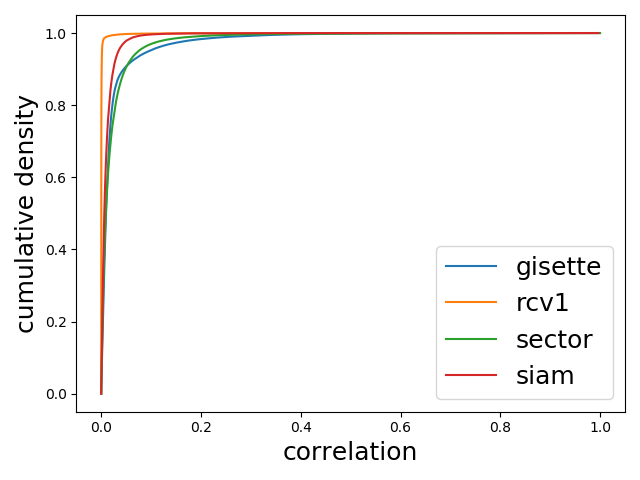}
\caption{Distribution of correlations. For each point $(x,y)$ on the line, $y$ is the empirical proportion of $(|correlation| \leq x)$.
Most of the correlations are close to zero, and only a few of them are significantly larger than zero.
}
\label{fig:cor_sparse}
\end{figure}

Covariance matrices of high dimensional datasets ~\cite{rutimann2009high, bien2011sparse, fan2016overview} are often sparse. For example, sparsity arises naturally in genomic datasets, where the genes from the same pathway are more likely to be closely correlated~\cite{pita2018pathway, luo2019batch} than different pathways. Figure~\ref{fig:cor_sparse} plots the distributions of the correlations of four high dimensional datasets (datasets taken from \cite{chang2011libsvm}). As we can see, most of the correlations are close to zero, and only a few of them are significantly larger than zero.

Assuming a sparse covariance matrix, our goal is to locate the large entries in the covariance matrix. Specifically, let $\mY = (Y_1, Y_2, \ldots, Y_d) \in \reals^{d}$ be a random vector of $d$ variables drawn from a joint distribution $F_{\mY}$ with sparse covariance matrix. We observe i.i.d. samples of the random vector $\mY$ sequentially, and at each time $t$, we can only access the sample $\mY^{(t)} \in \reals^d$. Our goal is to identify the large covariance pairs, i.e., pairs $(Y_i, Y_j)$  that satisfy $Cov(Y_i, Y_j) \geq u$ or $Cor(Y_i, Y_j) \geq u$, where $u\in \reals$ is some threshold separating large covariances from small ones. 

\textbf{Traditional covariance matrix sketching methods:}
Previous covariance matrix sketching methods rely on capturing the low-rank structure of the covariance matrix and compress the covariance matrix using random projections~\cite{bahmani2015sketching, chen2014robust}. However, though the projection is memory efficient and fast, the recovery step is expensive both memory-wise and computationally.
Also, the low-rank approximation may not apply for many relevant datasets. For example, when modeling gene expression, it is difficult to find small groups of genes whose expression levels dominate all other genes. Hence, it is worthwhile to consider covariance matrices with alternative structural assumptions. In this paper, we assume sparsity in covariance matrix for high dimensional data.

\textbf{Compressed sensing methods:}
Assuming that the covariance matrix is sparse, the compressed sensing~\cite{bahmani2015sketching} framework offers a natural solution for identifying large covariance pairs. However, most popular compressed sensing recovery procedures involve optimization over matrices of the original size, which is infeasible for large covariance matrices.

Count sketch (CS)~\cite{charikar2002finding} is a popular algorithm with a very efficient recovery procedure. CS can be used to store the empirical covariance matrix in sub-linear memory and provide an estimate for each covariance entry. When a new sample arrives, it is possible to calculate the update of the empirical covariance entries and add it to the sketch. After processing all the samples, we can select the top few signal covariances using the estimates of covariance entries from the count sketch.

However, performance of CS is poor when the signal to noise ratio (SNR) in data being inserted is relatively low. Count Sketch is a hashing based algorithm where collisions cause estimation errors. When the signal is substantially more significant than the noise (i.e., the SNR is large), count sketches can accurately locate the signals.~\cite{charikar2002finding}. However, in many practical settings, it is difficult to ensure that the empirical covariances of signal pairs are larger than those of noise pairs. For illustration, assume that the vectors $\mY$ follow a multivariate Gaussian distribution. Even if the covariance of a pair or variables $(i,j)$ is zero, the empirical covariance between $i$ and $j$ may still be large because of a large variance in the empirical covariance. The SNR is often not large enough to guarantee accurate recovery of signal pairs from count sketches. Not surprisingly, we observe relatively poor empirical performance of count sketches for sparse covariance matrix estimation.

In this paper, we focus on addressing this performance issue with count sketches when ingesting data with low SNR. 
We design a strategy that adds more samples of signal covariances while restricting the addition of noise from low covariances. 
This approach enhances the SNR of the data finally ingested by CS compared to a vanilla CS which inserts all the variables into the sketch.

\textbf{Central idea of Active Sampling Count Sketch (ASCS):}
Reducing hashing collisions is a critical step to increase the accuracy of identifying signal pairs. However, given that we have limited control over the hashing procedure, we chose to selectively filter the information added to the sketch to reduce the effect of collisions. We propose ASCS to reduce the noise samples inserted into the sketch. 
ASCS first invests some samples in exploring the magnitude of the large covariances (or signals). After this exploration, we compute a suitable, dynamically changing threshold that discriminates signal from noise. In the subsequent steps, when a new sample arrives, instead of inserting all the covariance pairs into the sketch, which increases the hashing collision errors, we only insert the covariance pairs with estimates above the threshold. Hence, ASCS reduces the hashing collision errors and improves the accuracy of detecting the signal covariances.

\textbf{Our contributions:}
We propose an algorithm for sparse covariance matrix estimation along with theoretical guarantees. 
Our experiments suggest that ASCS can provide a sparse estimation of a trillion scale covariance matrix efficiently in terms of both memory and computation. For example, the correlation matrix of the DNA $k$-mer dataset~\cite{vervier2016large} has 144 trillion unique entries, and storing the whole correlation matrix costs \textbf{0.5PB} (petabyte) memory. ASCS achieves $10^6\times$ compression and only uses \textbf{2GB} memory. It locates the top thousand correlation pairs with a mean correlation \textbf{close to one}, which is much better than that identified by vanilla CS (mean correlation $\bm{=0.35}$) at the same memory usage. To achieve comparable performance, CS requires \textbf{tenfold} memory.

Moreover, the practicality of our algorithm is evident from the running times. Our algorithm's running time is only governed by time taken to sketch the matrix. Even a naive implementation of ASCS can process and estimate matrix for ``url'' dataset~\cite{chang2011libsvm} (1.5GB, covariance matrix size: \textbf{20TB})  in less than \textbf{25 minutes} using \textbf{20MB} memory and DNA $k$-mer dataset (64GB, covariance matrix size: 0.5PB) in about 12-15 hours.
To the best of our knowledge, this paper shows a remarkable improvement in our ability to scale to large covariance matrices over traditional matrix sketching methods (like CS) and existing compressed sensing algorithms.

\vspace{-0.2cm}
\section{Related Work}

In this section we focus on literature that is relevant to our discussion in two aspects : (1) large scale matrix estimation and (2) adaptations of count sketch for improving prediction on top frequent elements. 

\textbf{Large scale matrix estimation:} 
\citet{pagh2013compressed} uses count sketch (AMS Sketch) to compute the matrix outer product when the product is sparse. To speeds up the computation, they first ``compress'' the matrix product into a polynomial expression. Then, they use FFT (Fast Fourier Transform) for polynomial multiplication.
When the matrices are sparse, the time complexity reduces from $O(d^2)$ to $O(N)$ ($N$ is the number of non-zero entries). \citet{pagh2013compressed} can also be used to compute the empirical covariance matrix in sub-quadratic time since a covariance matrix can also be expressed in the form an outer product.

\citet{valiant2015finding} proposed an sub-quadratic time complexity algorithm to search the large correlation pairs of Boolean vectors from a large number of extremely weakly correlated vectors. This algorithm is based on count sketches. To reduce the search complexity, vectors are randomly aggregated into sets of vectors (number of vector sets $\ll$ number of vectors). If a pair of vectors are highly correlated, we can still locate it by checking the correlations between the corresponding vector sets. Then, the algorithm uses brute force to search large correlation pairs among the vector sets that are closely correlated, while the searching time is significantly reduced since the number of vector sets $\ll$ number of vectors.

\citet{cormode2017fast} also target to find the large correlation pairs efficiently using the AMS sketch. The algorithm uses an AMS sketch to store the features one by one. 
Moreover, to encourage sub-quadratic computation complexity ($o(d^2)$), they encode the $d^2$ correlation pairs into $\Pi^2$ pairs of groups ($\Pi \ll d$). They also designed a decoder to recover the identities of large correlation entries in a particular group pair. However, to implement the algorithm, \citet{cormode2017fast} requires to have access to all the samples of a feature, which does not fit our setup since we assume the samples are observed sequentially.

\textbf{Adaptations of count sketch to improve prediction on frequent elements:}
Augmented Sketch~\cite{roy2016augmented} maintains a filter to count the high frequency queries outside of the sketch, which can reduce the hashing collisions of the high frequency queries. Cold Filter~\cite{zhou2018cold} proposed a solution similar to Augmented Sketch, where they use a sketch to filter out the low frequency items, and forward the high frequency query to another filter to get a better estimate of its frequency. In the section \ref{sec:smallexp}, we compare ASCS against these methods.

\vspace{-0.2cm}
\section{Problem statement}\label{sec2.1} \label{sec:probstatement}



Let $\mX = (X_1, X_2, \ldots, X_p) \in \reals^{p}$ be a vector following a distribution $F_\mX$. Let its expectation $\vmu = \EX{\mX}$ be $\alpha$-sparse, i.e., an entry picked uniformly at random is non-zero with probability $\alpha$,  $\PR{\mu_i \neq 0} =\alpha$. Let $\mX^{(1)}, \mX^{(2)}, \ldots, \mX^{(T)} \sim F_\mX$ be i.i.d samples observed sequentially, i.e., at time $t$, we can only access $\mX^{(t)}$. We refer to variables with zero mean as noise and others as signals. The goal in `Online Sparse Mean Estimation' is to identify the signals based on the observations $\mX^{(1)}, \mX^{(2)}, \ldots, \mX^{(T)}$.

Note that we assume that the observed samples are i.i.d distributed over time. This assumption is critical to the success of our algorithm. In real-world applications, we can induce randomness by buffering the incoming data and shuffling it before passing to the algorithm. This is a standard procedure and is used for large datasets in dataloaders of pytorch and tensorflow \cite{ketkar2017introduction, abadi2016tensorflow} for example.


\textbf{Online Sparse Covariance Estimation:}
This is a good model for estimating the non-zeros of a sparse covariance matrix. To see this, let $\mY = (Y_1,Y_2, \ldots, Y_d) \in \mb R^{d}$ be a random vector with distribution $F_\mY$ and with covariance matrix $\Sigma \in \reals^{d \times d}$.
Let $\mY^{(1)}, \mY^{(2)}, \ldots, \mY^{(T)} \sim F_\mY $ be i.i.d samples observed sequentially. Let $\mX \in \reals^{p}$ be the vector encoding the off-diagonal covariance entries of $\mY$, i.e., $\mX = \{Y_i Y_j - \EX{Y_i}\EX{Y_j}: 1 \leq i < j \leq d\}$ and $p = d(d-1)/2$. Let $\vmu = \EX{\mX}$.  Then the non-zero entries of the mean $\vmu$ correspond to the non-zeros of the covariance matrix.

\textbf{Notation:} 
We define the signal-to-noise (SNR) ratio of a data vector $\mX$ as follows. Let $\mX_S$ and $\mX_N$ be the data vector containing signal and noise variables of $\mX$ respectively. Then, the $SNR(\mX)$ is defined as 
\begin{align*}
    SNR(\mX) = \EX{\norm{\mX_S}_2^2}/\EX{\norm{\mX_N}_2^2}
\end{align*}
We can similarly define the signal to noise ratio of the data vector at time $t$.

\textbf{Performance metric:} We evaluate the performance of an algorithm on this problem with two metrics (1) If $\mS \subseteq \mX$ is the set of variables with highest estimated mean ($\hat{\vmu}$) as reported by the algorithm. Then the first metric we use is $\textrm{avg} \{ \vmu_i | X_i \in \mS \}$ , i.e. average of true means of the variables. (2) The accuracy of the algorithm to classify the variables as signal $(\vmu_i \neq 0)$ or noise $(\vmu_i = 0)$

\vspace{-0.2cm}
\section{Na\"ive Proposal: Count Sketch} \label{sec:naivecs}


We start by describing how count sketches can in principle be used to concisely locate and recover the signals of heavy hitters of $\vmu$.
Count sketch is a probabilistic data structure widely used to identify heavy hitters in streaming data. A count sketch consists of $K$ hash tables and each hash table has $R$ buckets. The streaming data are mapped into the hash tables using $K$ independent hash functions. Let $\mW \in \reals^{K \times R}$ be the matrix storing the values in the count sketches, and let $h_1, h_2, \ldots, h_K$ be independent uniform hash functions $h_e\colon \{1,2,\ldots,p\} \to \{1,2,\ldots,R\}$. In addition, count sketch uses sign hash functions $s_e\colon \{1,2,\ldots,p\} \to \{+1, -1\}$ to map the components of the vectors randomly to $\{+1, -1\}$.
To estimate $\mu_i$, we need to use the count sketch to store the sample mean of $X_i$, $\frac{1}{T}\sum_{t=1}^{T} X^{(t)}_i$. When a new sample of $X_i$, $X^{(t)}_i$ arrives, count sketch calculates the hash locations $h_e(i)$, $e = 1, \ldots, K$, and adds $\left(\frac{1}{T}X^{(t)}_i\right)\cdot s_e(i)$ to the corresponding bucket (Algorithm~\ref{alg:algo1}, line 3 to 5). In order to retrieve the estimated value of $\mu_i$, count sketch computes the hashing locations, $(h_1(i), h_2(i), \ldots,h_K(i))$ and retrieves the values stored in the corresponding buckets. Then, take the median of the retrieved values as the estimate, $\hat{\mu}_i = \median_e \mW_{e, h_e(i)} \cdot s_e(i)$.  

\vspace{-0.3cm}
\begin{algorithm}[H]
\caption{Count Sketch Algorithm}
\label{alg:algo1}
\begin{algorithmic}[1]
  \STATE \textbf{Input}: $\mX^{(t)}$ for $t=1,2,\ldots,T$,  $K$ independent uniform hash functions $h_1, h_2,\ldots, h_K$, and sign functions $s_1, s_2, \ldots, s_K$
  \STATE Initialize entries of hash table array $\mW \in \reals^{K \times R}$ to zero
  \FOR{$t = 1,2,\ldots, T$}
  \STATE Insertion: update $\mW_{e, h_e(i)} += \frac{1}{T}X^{(t)}_i \cdot s_e(i)$ for $e=1,2,\ldots,K$ and $i = 1,2,\ldots, p$
  \ENDFOR
  \STATE \textbf{Retrieval:} estimate of $\mu_i$, $\hat{\mu}_i = \median_e \mW_{e, h_e(i)} \cdot s_e(i)$
\end{algorithmic}
\end{algorithm}
\vspace{-0.3cm}


\textbf{Updates of empirical covariance entries:} 
First, we define the following notations to represent the sample means.
\begin{align*}
    \bar{\mX} = \frac{1}{T} \sum_{i=1}^T \mX^{(i)}, \,  \bar{\mY} = \frac{1}{T} \sum_{i=1}^T \mY ^{(i)}, \,
    \bar{\mX}^{(t)} = \frac{1}{t} \sum_{i=1}^t \mX^{(i)}, \, \bar{\mY}^{(t)} = \frac{1}{t} \sum_{i=1}^t \mY^{(i)}
\end{align*}
Let $X_i$ correspond to the covariance between $Y_a$ and $Y_b$.  Theoretically, $X^{(t)}_i = (Y^{(t)}_a - \EX{Y_a}) (Y^{(t)}_b - \EX{Y_b})$. However, $\EX{Y_a}$ and $\EX{Y_b}$ are not available in practice. Hence, at time $t$, we use the sample mean $\bar{\mY}^{(t)}$ to replace $\EX{\mY}$. Hence, to implement count sketch, we should maintain a vector storing the $\bar{\mY}^{(t)}$ (this vector is updated as new samples are processed).
Then, to use algorithm~\ref{alg:algo1} to store the empirical covariances in the count sketch, we need to ensure $\sum_{k=1}^{t} X^{(k)}_i = \sum_{k=1}^{t} (Y^{(k)}_a - \bar{Y}^{(t)}_a) (Y^{(k)}_b - \bar{Y}^{(t)}_b)$ for all $t = 1,2,\ldots,T$. To update the count sketch from time $t$ to $t+1$, we need to calculate $X^{(t+1)}_i$ using the new sample $\mY^{(t+1)}$ and empirical mean of $\mY$ stored in the vector. $X^{(t+1)}_i$ can be expressed as $X^{(t+1)}_i = (Y^{(t+1)}_a - \bar{Y}^{(t+1)}_a) (Y^{(t+1)}_b - \bar{Y}^{(t+1)}_b) + adjustment$. The adjustment comes from the changes of empirical mean from $\bar{\mY}^{(t)}$ to $\bar{\mY}^{(t+1)}$ for previously inserted samples. Moreover, the ``adjustment'' term only depends on the current new sample $\mY^{(t+1)}$ and $\bar{\mY}^{(t)}$ ($\bar{\mY}^{(t+1)}$ is computed from $\mY^{(t+1)}$ and $\bar{\mY}^{(t)}$). 
\begin{align}
adjustment
=& (t+1)(\bar{Y}^{(t)}_a - \bar{Y}^{(t+1)}_a)(\bar{Y}^{(t)}_b - \bar{Y}^{(t+1)}_b) + \nonumber  \\
& (Y^{(t+1)}_a - \bar{Y}^{(t)}_a)(\bar{Y}^{(t)}_b - \bar{Y}^{(t+1)}_b)  + \nonumber  \\
& (\bar{Y}^{(t)}_a - \bar{Y}^{(t+1)}_a)(Y^{(t+1)}_b - \bar{Y}^{(t)}_b)   \nonumber 
\end{align}

In the real experiments, when $t$ is large enough, the ``adjustment'' is very small and almost negligible. We may just skip the ``adjustment'' term to speed up the computation.

\vspace{-0.2cm}
\subsection{Shortcomings of Count Sketch with low SNR data}

The count sketch algorithm is guaranteed to locate signals with high probability when the empirical mean of signal variables are much larger than those of noise variables. The estimate of the means $\mu_i$ from the count sketch is
\vspace{-0.2cm}
\begin{align}
\hat{\mu}_i 
&= \median_e \mW_{e, h_e(i)}\cdot s_e(i)  \\
&= \bar{X}_i + \median_e \sum_{j:h_e(j) =h_e(i)} \bar{X}_j \cdot I(i \neq j) \cdot s_e(i)   \nonumber \\
&= \bar{X}_i + \median_e H_e(i),   \nonumber
\end{align}
where $\bar{X}_i$ is the sample mean of $X_i$, and $H_e(i) = \sum_{j:h_e(j)=h_e(i)} \bar{X}_j \cdot I(i \neq j) \cdot s_e(i)$. Here, $H_e(i)$ is the noise in the $X_i$ estimate due to collisions in the $e$-th hash table.

Due to the sparsity of signals, $\median_e H_e(i)$ only contains noise variables with high probability. If the sample mean of a signal variable $\bar{X}_i$ is much larger than that of noise variables we would still have $\abs{\bar{X}_i} \gg \abs{H_e(i)}$  even though $H_e(i)$ aggregates some noise variables. In such cases $\hat{\mu}_i$ is highly discriminating between signals and noises.
However, due to the potentially large variance of sample covariances, the sample mean of some noise variables may be close to that of the signals. In this case, hashing collision would be too large, and $\hat{\mu}_i$ would not be able to differentiate signals from noises with high accuracy.


\vspace{-0.2cm}
\section{Our Proposal: Active Sampling Count Sketch (ASCS)} \label{sec:ascs}

To accurately locate the signal variables, it is critical to reduce the size of noise variables and raise the signal-to-noise ratio. We leverage active sampling to achieve this goal. Since the estimation error using CS arises from the hashing collisions, and most of the collision variables are noise variables (due to the sparsity of signal variables), if we can reduce the noise variable samples inserted into the count sketch, the size of hashing collisions can be reduced.

ASCS includes two stages, an exploration period and a sampling period. During the exploration period, we insert all the variables into the count sketch to obtain a coarse estimate of $\mu_i$ (Algorithm~\ref{alg:algo2}, line 4-7). After the exploration period, we only sample the variables whose estimates are larger than a predetermined threshold $\tau^{(t)}$ at each round (Algorithm~\ref{alg:algo2}, line 9-13). Hence, in contrast to the vanilla count sketch, where all the variables are inserted into the sketch, ASCS only samples the variables with large estimates. Therefore, the hashing collision scale is reduced.

\vspace{-0.3cm}
\begin{algorithm}[H]
\caption{Active Sampling Count Sketch Algorithm}
\label{alg:algo2}
\begin{algorithmic}[1]
  \STATE \textbf{Input}: $\mX^{(t)}$ for $t=1,2,\ldots,T$,  $K$ independent uniform hash functions $h_1, h_2,\ldots, h_K$, and sign functions $s_1, s_2, \ldots, s_K$
  \STATE \textbf{Parameters}: search the length of exploration period $T_0$ and sampling thresholds $\tau^{(t)}$ using Algorithm 3
  \STATE Initialize entries of hash table array $\mW \in \reals^{K \times R}$ to zero,
  
  \STATE \#\# Exploration Period
  \FOR{$t = 1,2,\ldots, T_0$}
  \STATE Insertion: update $\mW_{e, h_e(i)} += \frac{1}{T} X^{(t)}_i \cdot s_e(i)$ for $e=1,2,\ldots,K$ and $i = 1,2,\ldots, p$
  \ENDFOR
  
  \STATE \#\# Sampling Period
  \FOR{$t = T_0+1,T_0+2,\ldots, T$}
  \STATE Retrieve estimate of $\mu_i$, $\hat{\mu}^{(t-1)}_i = \median_e \mW_{e, h_e(i)} \cdot s_e(i)$
  \IF {$\hat{\mu}^{(t-1)}_i \geq \tau^{(t-1)}$}
  \STATE Insertion: update $\mW_{e, h_e(i)} += \frac{1}{T} X^{(t)}_i \cdot s_e(i)$ for $e=1,2,\ldots,K$ and $i = 1,2,\ldots, p$
  \ENDIF
  \ENDFOR
  
  \STATE Retrieval: estimate of $\mu_i$, $\hat{\mu}^{(T)}_i = \median_e \mW_{e, h_e(i)} \cdot s_e(i)$
\end{algorithmic}
\end{algorithm}
\vspace{-0.3cm}

\begin{figure}[ht]
\centering 
\includegraphics[width=0.3 \textwidth]{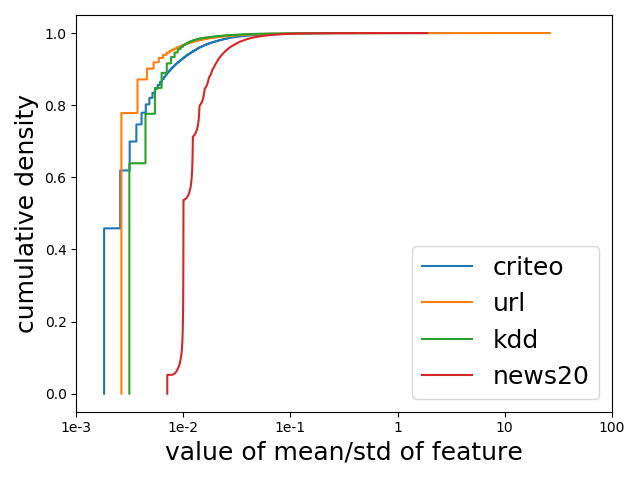}
\caption{Distribution of "mean/std" values for features. For each point $(x,y)$ on the line, $y$ is the empirical proportion of $(|mean/std| \leq x)$. The mean of most of features are almost negligible compared to the std.}
\label{fig:mean_sparse}
\vspace{-0.3cm}
\end{figure}

\textbf{Computational cost of CS and ASCS in high-dimensional sparse datasets:}
Sparsity is commonly observed in most high dimensional datasets. Moreover, most of the features also have very small means with respect to the standard deviation (i.e. normalized by standard deviation to remove the scaling effects). Figure \ref{fig:mean_sparse} shows the distribution of $mean/std$ values for features of some high dimensional datasets (datasets taken from \cite{chang2011libsvm}). As we can see the mean of most of the features have extremely low (less than 1\% of its standard deviation). 

\vspace{-0.3cm}

Hence, for pairs ($Y_a, Y_b$), if any of the two features have mean/std value close to 0, we can ignore the rightmost term of equation~\eqref{eq_2} and approximate the covariance $Cov(Y_a, Y_b) \approx \EX{Y_a Y_b}$. 

\begin{equation}
\frac{Cov(Y_a, Y_b)}{\sigma(Y_a)\sigma(Y_b)} = \frac{\EX{Y_aY_b}}{\sigma(Y_a)\sigma(Y_b)} - \frac{\EX{Y_a}}{\sigma(Y_a)}\frac{\EX{Y_b}}{\sigma(Y_b)} \label{eq_2} 
\end{equation}
\vspace{-0.2cm}

Then, during insertion stage (Algorithm~\ref{alg:algo2} line 6), we just need to calculate and insert $\frac{1}{T} Y^{(t)}_a Y^{(t)}_b$ into the hash tables. Moreover, if we observe the value of a feature at time $t$ equals to 0, i.e.$Y^{(t)}_a = 0$. Then, $\frac{1}{T} Y^{(t)}_a Y^{(t)}_b = 0$ for all the $Y^{(t)}_b$, and we can skip these computation and insertion operations.

In the worst case, the computation cost of both CS and ASCS is $O(Td^2)$ where $T$ is the number of samples and $d$ is the dimension of data features. For the high dimensional sparse datasets, with the approximations mentioned above, the computation complexity is reduced to $O(T(n_z+n_u)^2)$ where $n_z$ is the number of non-zeros per sample, and $n_u$ is the number of features whose means are not approximated to zero. As most high dimensional datasets are sparse ($n_z$ is small) and as shown in figure \ref{fig:mean_sparse}, $n_u$ is generally observed to be small in such datasets, we can efficiently run ASCS on most high dimensional datasets.

\vspace{-0.2cm}
\section{Choice of hyperparameters}

In this section, we discuss how the parameters of the ASCS algorithm, including the length of the exploration period $T_0$ and the sampling thresholds at each time step, $\tau^{(t)}$, can be chosen in a principled manner. 
To analyse ASCS parameters, we need to make additional assumptions on distribution of $X_i$. These assumptions are not strictly necessary for the algorithm to work but are made simplify the analysis.

This section is organized as follows. We first introduce distribution assumptions in sections~\ref{sec:assumption} . Next, we validate the assumptions empirically on the simulation and real-world datasets in section~\ref{sec:assumption_validation}. In sections~\ref{sec:algorithm3}, \ref{sec:exploration} and~\ref{sec:tau}, we describe the algorithms to choose hyperparameters and provide theoretical guarantees on performance of ASCS under this choice.

\vspace{-0.2cm}
\subsection{Distribution assumptions and their motivation}\label{sec:assumption}\label{sec_4.4}

We make two assumptions on $X_i$ to simplify the analysis of ASCS. As shown in subsequent subsections, these assumptions are well supported empirically on both simulation and real data.
\begin{itemize}[leftmargin=*,nosep]
    \item \textbf{Independence assumption:} We ignore the dependence between $X_i$s, and assume that $X_i$s are independent to each other. 
    \paragraph{Motivation}
    For a sparse covariance matrix, the $X_i$ are approximately independent. 
    For example, if the empirical covariance matrix follows a Wishart distribution, which is a common model for the empirical covariances, most of the covariance entries are independent to each other when the true covariance matrix is sparse~\cite{christensen2015covariance}. Even though the $X_i$ are not perfectly independent, the experiments in the following section show that the independence assumption is a good approximation for simulated and real datasets.
    
    \item \textbf{Gaussian distribution assumption of $\bar{X}^{(t)}_i$:} We assume that the empirical average of $X_i$, $\bar{X}^{(t)}_i \sim N(\mu_i, \sigma^2/t)$ where $\gamma < t \leq T$ ($\gamma$ is a relatively large constant to guarantee the validity of central limit theorem). Also, we assume $\mu_i=u>0$ if $\mu_i \neq 0$.
    \paragraph{Motivation}
    We know by central limit theorem that normal distribution is a good assumption on sample mean $\bar{X}^{(t)}_i$ when $t$ is sufficiently large ($t > \gamma$) provided the tail probability of $X_i$ is controlled. 
    If the $X_i$ are sub-Gaussian, then $\bar{X}^{(t)}_i$ behaves like a Gaussian random variable for all $t$. Even if the $X_i$ have a heavier tail than a sub-Gaussian random variable, this sum concentrates around its mean and we can control its deviation from the mean. 
    So, the distribution of $\bar{X}^{(t)}_i$ only deviates from a Gaussian distribution in the region that is very far away from $\mu_i$ (see Figure~\ref{fig:norm_assump}). Moreover, most of the high dimensional datasets have thousands or millions of samples, thus $t$ is large enough. Therefore, to simplify the proof of theorems, we simply assume that $\bar{X}^{(t)}_i$ follows a Gaussian distribution for any $t > \gamma$.
\end{itemize}

\vspace{-0.2cm}
\subsection{Empirical justification of the distribution assumptions} \label{sec:assumption_validation}
\label{sec_4.5}

In this section, we empirically demonstrate that our distributional assumptions are approximately satisfied for the distribution of covariance entries. 

\vspace{-0.5cm}
\begin{figure}[ht]
\centering 
\begin{subfigure}{0.499\linewidth}
  \includegraphics[trim=0 0 0 0,clip, width=\textwidth]{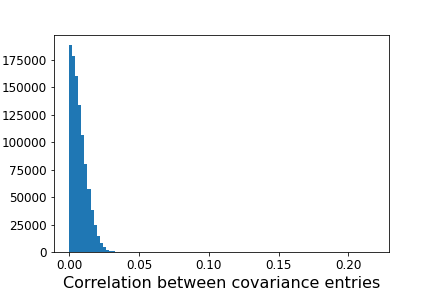}
\end{subfigure}\hfil
\begin{subfigure}{0.499\linewidth}
  \includegraphics[trim=0 0 0 0,clip,width=\textwidth]{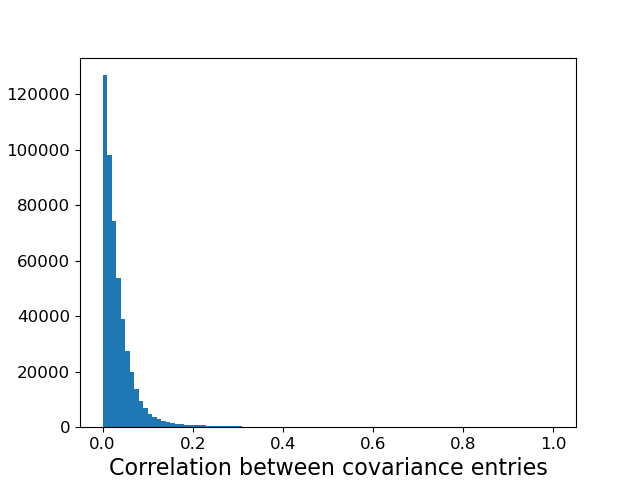}\hfil
\end{subfigure}
\caption{Histogram of the absolute correlations between the covariance entries. (Left) simulation dataset (Right) "gisette" dataset. Most of the correlations are very close to zero.}
\label{fig:cor_of_cov}
\end{figure}
\vspace{-0.5cm}

We perform the experiments on two sets of data. (1) simulation data (2)  "gisette": real data which is used in our experiments later. 
For simulation dataset, we simulate multiple normal datasets using a true covariance matrix where we set the proportion of signal covariance to $\alpha = 0.5\%$ (the others covariates are noise covariances and are set to 0). The strength of signal covariances are uniformly sampled between $0.5$ and $1$. All datasets follow the same distribution.  As``gisette'' has limited number of samples ($6,000$), we sample similar datasets from original data using bootstrapping. For both the simulation and ``gisette'' cases, we sample 15,000 datasets respectively where each dataset contains 1,000 features and 1,000 samples ($T=1000, d=1000, p=(d-1)d/2$).

 For each of these datasets, we compute the empirical covariances using the first 150 samples, $\bar{X}^{(t)}_i$, where $t=150$. Since we simulated 15,000 datasets, for each empirical covariance entry, we collect 15,000 independent samples.


\begin{figure*}[ht]
\centering 
\begin{subfigure}{0.25\textwidth}
  \includegraphics[trim=0 0 0 0,clip,width=\linewidth]{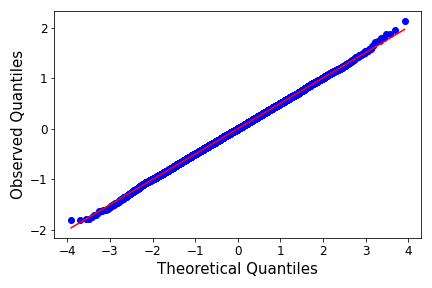}
\end{subfigure}\hfil 
\begin{subfigure}{0.25\textwidth}
  \includegraphics[trim=0 0 0 0,clip,width=\linewidth]{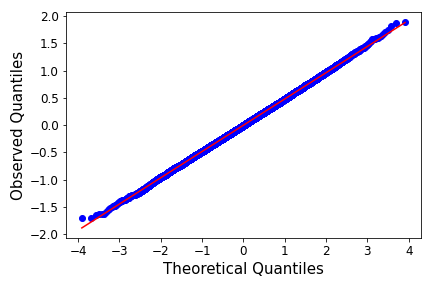}
\end{subfigure}\hfil 
\begin{subfigure}{0.25\textwidth}
  \includegraphics[trim=0 0 0 0,clip,width=\linewidth]{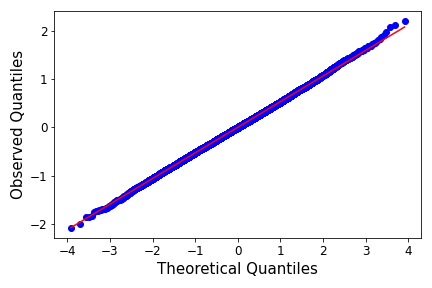}
\end{subfigure}\hfil
\begin{subfigure}{0.25\textwidth}
  \includegraphics[trim=0 0 0 0,clip,width=\linewidth]{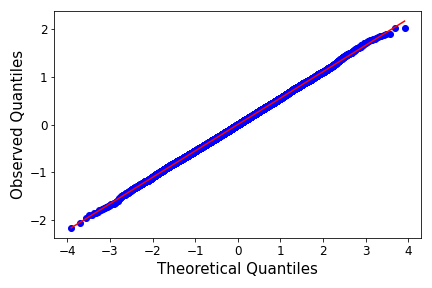}
\end{subfigure}

\begin{subfigure}{0.25\textwidth}
  \includegraphics[trim=0 0 0 0,clip,width=\linewidth]{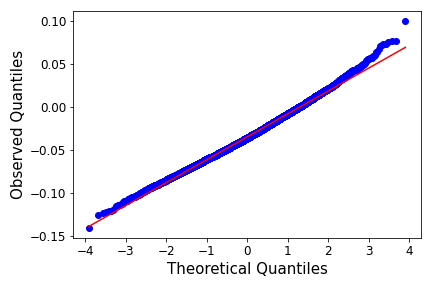}
\end{subfigure}\hfil 
\begin{subfigure}{0.25\textwidth}
  \includegraphics[trim=0 0 0 0,clip,width=\linewidth]{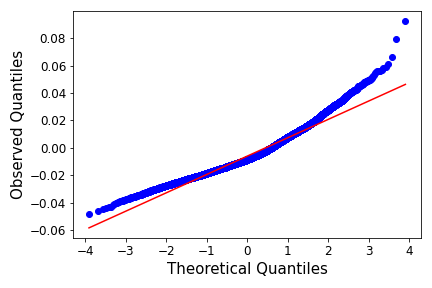}
\end{subfigure}\hfil 
\begin{subfigure}{0.25\textwidth}
  \includegraphics[trim=0 0 0 0,clip,width=\linewidth]{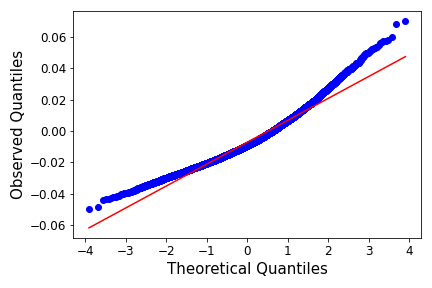}
\end{subfigure}\hfil
\begin{subfigure}{0.25\textwidth}
  \includegraphics[trim=0 0 0 0,clip,width=\linewidth]{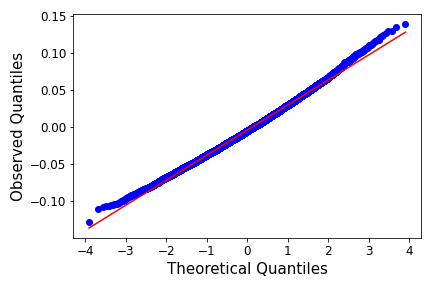}
\end{subfigure}
\caption{(Top) The QQ-plots compares the distribution of four randomly selected covariance entries to the normal distribution for simulation dataset; (Bottom) QQ-plots for some covariance entries from "gisette" dataset. The distribution of an empirical covariance can be well approximated by a Gaussian distribution.}
\label{fig:norm_assump}
\vspace{-0.4cm}
\end{figure*}

\textbf{Independence assumption:} We evaluate the linear independence of $(X_i,X_j)$ pairs by comparing the correlation for the pair $(\bar{X}_i^{(t)},\bar{X}_j^{(t)})$. As declared in the previous section, the covariances are not perfectly independent. However, figure~\ref{fig:cor_of_cov} suggests that most of the covariances are very weakly correlated. For example, on the simulation dataset, over 97\% of the covariances pairs having correlations less than 0.02. Hence, assuming the covariances are independent is reasonable and close to the situation in the real datasets.

\textbf{Gaussian distribution assumption:} We assumed $\bar{X}^{(t)}_i$ following a normal distribution, i.e. if $X_i$ corresponds to a noise covariance entry, $\bar{X}^{(t)}_i \sim N(0, \sigma^2/t)$. Here, we set $t=150$. Figure~\ref{fig:norm_assump} randomly selects four covariance entries and compares the marginal distribution of the empirical covariances to a normal distribution. Obviously, the normal distribution assumption fits perfectly on the simulation dataset. On the ``gisette'' dataset, though the distribution of  $\bar{X}^{(t)}_i$ is slightly skewed for some $X_i$s (in Figure~\ref{fig:norm_assump}, right skewed), normal distribution assumption is still a good approximation. For high dimensional large datasets, usually $t$ is much larger since we have a lot of samples. And the distribution of $\bar{X}^{(t)}_i$ is even closer to a Gaussian distribution as $t$ increases. 

\vspace{-0.2cm}
\subsection{Algorithm for determining the hyperparameters of ASCS} \label{sec:algorithm3}

We next state algorithm~\ref{alg:algo3} for determining the hyperparameters of ASCS, including the length of the exploration period $T_0$ and the sampling thresholds $\tau^{(t)}$. 

ASCS incorporates an active sampling procedure with count sketch, which comes at a cost because it may miss some signal variables and leads to a reduction of the signal strength. Therefore, algorithm~\ref{alg:algo3} ensures that the sampled variables at each round cover most of the signal variables while at the same time, filtering enough noise variables. First, algorithm~\ref{alg:algo3} determines the length of exploration period $T_0$, such that the probability of missing a signal variable at time $T_0$ (at the beginning of sampling) is upper bounded $\delta$ (line 2). Then, algorithm~\ref{alg:algo3} chooses a $\theta$ such that after time $T_0$, the ASCS will miss a signal variable with probability at most $\delta^* - \delta$ (line 3). Since $\tau^{(t)}$ is a linear function of $\theta$, $\tau^{(t)}$ is determined once we find a suitable $\theta$. Combining line 2 and line 3, algorithm~\ref{alg:algo3} can determine the hyperparameters of ASCS that control the probability of missing a signal variable throughout the sampling procedure (the probability is upper bounded by $\delta^*$). We expand the details in the following two subsections.

\subsection{Hyperparameter: length of exploration period ($T_0$)} \label{sec:exploration}
\label{sec4.1}

ASCS includes a sampling procedure to sample the variables with large $\mu_i$. But before sampling the variables, we need an exploration period ($T_0$) to explore the values of $\mu_i$. 
\begin{itemize}
    \item If $T_0$ is too large, the sampling period ($T_0, T$] is not long enough to filter enough noise variables. 
    \item If $T_0$ is too small, ASCS may wrongly filter too many signal variables. 
\end{itemize}

Therefore, we \textbf{choose a $T_0$ as small as possible while sufficiently large to control the risk of missing signals during the first sampling operation}. Specifically, we require that signal variables are missed with small probability, i.e. $\PR{\abs{\hat{\mu}^{(T_0)}_i} < \tau^{(T_0)} \mid \mu_i = u} \leq \delta$ where $\delta \in (0,1)$ is an input hyperparameter and usually takes a small positive value. Moreover, to start sampling as early as possible, we assign a small positive value for $\tau^{(T_0)}$. \textbf{The following theorem upper bounds the probability of missing a signal covariance at $T_0$:}

\vspace{-0.4cm}
\begin{algorithm}[ht]
\caption{Determine $T_0$ and $\tau^{(t)}$ of ASCS}
\label{alg:algo3}
\begin{algorithmic}[1]
    \STATE \textbf{Input}: Initial probability of missing a signal variable $\delta$ and the total probability of missing a signal variable $\delta^*$; initial sampling threshold $\tau^{(T_0)}$ 
    \STATE \textbf{Find $T_0$}: minimum $T_0$ satisfying $\Phi\left(-\frac{\sqrt{t}u-\frac{T\tau^{(t)}}{\sqrt{t}}}{\kappa\sigma}\right) p_0^K + (1-p_0^K) \leq \delta$ (ensure LHS of equation~\eqref{eq1} $\leq \delta$). 
    \STATE \textbf{Find $\tau^{(t)}$}: 
    \begin{enumerate}
        \item Set $\tau^{(t)} = \tau^{(T_0)} + \frac{\theta}{T}(t-T_0)$ 
        \item $\theta$: maximum $\theta^{'}$ satisfying \\ $\exp\left[\frac{(u-\theta^{'})(\tau^{(T_0)} - \frac{T_0}{T}\theta^{'})}{\omega^2_1} \right] \Phi\left(\frac{T_0(2\theta^{'}-u)-\tau^{(T_0)}T} {\sqrt{T_0} \omega_1}\right) \leq \delta^* - \delta$ (ensure LHS of equation~\eqref{eq2} $\leq \delta^* - \delta$)
    \end{enumerate}
\end{algorithmic}
\end{algorithm}
\vspace{-0.5cm}

\begin{theorem}
\label{theorem_1}
Let $\mX$ be the random vector defined in section~\ref{sec2.1} and let $\hat{\mu}^{(T_0)}_i$ to the estimate the mean of a covariance variable $X_i$ from the sketch at time $T_0$. When the ASCS only has one hash table (K=1), then following Algorithm~\ref{alg:algo2}, for any given $\gamma \leq T_0 \leq T$ and sampling threshold $\tau^{(T_0)} \in (0, u)$, the probability of missing a signal variable $X_i$ is upper bounded by,
\begin{align}
\PR{\abs{\hat{\mu}^{(T_0)}_i} < \tau^{(T_0)} \mid \mu_i = u}
\leq
\Phi\left(-\frac{\sqrt{T_0}u-\frac{T\tau^{(T_0)}}{\sqrt{T_0}}}{\kappa_0\sigma}\right) p_0 + (1-p_0),
\label{eq1}
\end{align}
where $\kappa_0 = \sqrt{1 + \frac{(p-1)(1-\alpha)}{R-\alpha}}$ and
$p_0 = \left(\frac{R-\alpha}{R}\right)^{p-1}$ are independent of $\tau^{(T_0)}$ and $T_0$. 
$\Phi(\cdot)$ is the c.d.f of the standard normal distribution. 
\end{theorem}

\textbf{Interpretation:}
With the probability bound in  theorem~\ref{theorem_1}, we can easily use binary search to find the minimum $T_0$ satisfying the requirement. Note that the RHS of equation~\eqref{eq1} is always larger than $1-p_0$, and we define $1-p_0$ as ``saturation probability'' (SP). And $\delta$ should be larger than $1-p_0$ to ensure a feasible $T_0$. 

Also note that theorem~\ref{theorem_1} requires ASCS only having one hash table, while most of the count sketch designs include multiple hash tables to achieve smaller estimation errors. 
We can also upper bound the probability of LHS of equation~\eqref{eq1} \textbf{for multiple hash tables} cases, where the LHS of equation~\eqref{eq1} is upper bounded by an expression close to the RHS of equation~\eqref{eq1}, with $\kappa_0$ replaced by $\kappa = \sqrt{1 + \frac{\pi(p-1)(1-\alpha)}{2K(R-\alpha)}}$, and $p_0$ replaced by $p^K_0$ (We did not provide the exact upper bound of LHS of equation~\eqref{eq1} since it does not have a closed form solution). The approximation comes from approximating the sample median of standard normal variables with its asymptotic distribution. 


\subsection{Hyperparameter: sampling thresholds ($\tau^{(t)}$)}\label{sec:tau}

After the exploration period, ASCS starts to sample variables that are added to the sketches. The sampling procedure requires us to design a sampling threshold $\tau^{(t)}$ at time $t$. Variables whose estimates are above $\tau^{(t)}$ will be added to the sketches.
\begin{itemize}
    \item If $\tau^{(t)}$ is too large, ASCS filters too many signal variables.
    \item If $\tau^{(t)}$ is too small, ASCS adds too many noise variables to the sketches, thus, decreases the signal-to-noise (SNR) ratio and reduces the accuracy of locating signal variables.   
\end{itemize}
Hence, we desire a $\tau^{(t)}$ as large as possible while sufficiently small to control the risk of a signal variable.

However, directly optimizing $\tau^{(t)}$ for all $T_0 \leq t < T$ is difficult. Instead, we restrict the choice of $\tau^{(t)}$ by raising $\tau^{(t)}$ linearly with $t$, i.e. $\tau^{(t)} = \tau^{(T_0)} + \frac{\theta}{T}(t-T_0)$, where $\theta$ is a non-negative parameter. We choose the linear model for $\tau^{(t)}$ since it only includes two parameters $\tau^{(T_0)}$ and $\theta$, which is simple enough to tune in practice. Moreover, to ensure ASCS won't miss too many signals, by law of iterated logarithm~\cite{kolmogoroff1929gesetz, jamieson2014lil}, $\tau^{(t)}$ should be raised at most $\mc O((t-T_0)\log(t-T_0))$, while linear model is close to optimal.  
With this design of $\tau^{(t)}$ we can leverage \textbf{the following theorem upper bounds probability of omitting a signal variable during the sampling procedure}. 


\begin{theorem}
\label{theorem_2}
Let the sampling threshold designed as $\tau^{(t)} = \tau^{(T_0)} + \frac{\theta}{T}(t-T_0)$ for $T_0 \leq t \leq T$. $I(i)$ is an indicator variable and $I(i)=0$ if the hashing collision of $X_i$ only includes noise variables. 
ASCS only has one hash table ($K=1$). Following Algorithm~\ref{alg:algo2}, if the signal variable $X_i$ only collides with noise variables ($I(i)=0$), then for any $0 < \theta < u$, the probability that a signal variable $X_i$ is omitted during sampling at any time $T_0 < t \leq T$ is upper bounded by,
\begin{align}
&\PR{\exists t \leq (T_0, T], \abs{\hat{\mu}^{(t)}_i} \leq \tau^{(t)}, \hat{\mu}^{(T_0)}_i > \tau^{(T_0)} \mid \mu_i=u, I(i)=0}     \nonumber  \\
\leq&
\exp\left[\frac{(u-\theta)(\tau^{(T_0)} - \frac{T_0}{T}\theta)}{\omega^2} \right] 
\Phi\left(\frac{T_0(2\theta-u)-\tau^{(T_0)}T} {\sqrt{T_0} \omega}\right),     
\label{eq2}
\end{align}
where $\omega$ is independent of the sampling parameters, \\
$\omega^2 
= \frac{1}{T_0}\Var{\hat{\mu}^{(T_0)}_i \mid I(i)=0} 
= \sigma^2 \left(1 + \frac{(p-1)(1-\alpha)} {T^2(R-\alpha)}\right)$.
\end{theorem}

\textbf{Interpretation:}
Theorem~\ref{theorem_2} upper bounds the probability that $X_i$ is filtered after $T_0$. Hence, combining with theorem~\ref{theorem_1}, we can upper bound the probability of omitting $X_i$ throughout the sampling procedure. Specifically, if we want ASCS to omit a signal variable with probability at most $\delta^*$ ($\delta^*$ is an input hyperparameter. A relatively large $\delta^*$ is preferred when the data SNR is low). We just need to choose a $\theta$ such that the RHS of \eqref{eq2} $= \delta^* - \delta$.


Note that theorem~\ref{theorem_2} only considers the case of $I(i)=0$. If $I(i) = 1$, meaning that $X_i$ collides with other signal variables, theorem~\ref{theorem_1} has assumed $X_i$ is always filtered by ASCS at time $T_0$ (which is the worst case).  Hence, theorem~\ref{theorem_2} does not need to consider it again.

Similar to theorem~\ref{theorem_1}, theorem~\ref{theorem_2} is restricted to ASCS only having one hash table. When ASCS \textbf{has multiple hash tables}, the LHS probability of equation~\eqref{eq2} is also upper bounded. The upper bound can be approximated similar to the RHS of equation~\eqref{eq2}, where the $\omega^2$ is replaced by $\omega^2_1$, and $\omega^2_1 
= \frac{1}{T_0}\Var{\hat{\mu}^{(T_0)}_i \mid I(i)=0} 
= \sigma^2 \left(1 + \frac{\pi(p-1)(1-\alpha)} {2KT^2(R-\alpha)}\right)$. Again, the approximation comes from approximating the sample median of standard normal variables.  

\vspace{-0.2cm}
\section{Analysis of ASCS}

In the previous section, we have introduced additional assumptions about $X_i$, and develop theoretical guarantees for ASCS using these assumptions. In the following subsection, we will show that ASCS can increase the signal-to-noise ratio (SNR) of the data before inserting into CS.

\vspace{-0.2cm}
\subsection{ASCS increases SNR}

Compared to na\"ive data ingestion in  count sketch, ASCS reduces the noise variables inserted into the sketch without losing too many signal variable. 
Let $\mX^{(t)}_S$ and $\mX^{(t)}_N$ contain the signal and noise variables of $\mX^{(t)}$ that ASCS adds to the sketch:
\begin{eqnarray}
\mX^{(t)}_S= \left\{X^{(t)}_i: \abs{\hat{\mu}^{(t-1)}_i} \geq \tau^{(t-1)} \ \text{and} \ \mu_i = u\right\},   \nonumber
\\
\mX^{(t)}_N = \left\{X^{(t)}_i: \abs{\hat{\mu}^{(t-1)}_i} \geq \tau^{(t-1)} \ \text{and} \ \mu_i = 0\right\}.  \nonumber
\end{eqnarray}

We define the signal-to-noise ratio of $t$-th sample as 
\begin{align*}
\text{SNR}^{(t)} 
= \EX{\norm[2]{\mX^{(t)}_S}^2} / \EX{\norm[2]{\mX^{(t)}_N}^2}    
\end{align*}
where the expectation is taken w.r.t the first $t$ random samples of $\mX$. 
For the vanilla CS, $\mX^{(t)}_S$ and $\mX^{(t)}_N$ include all the signal and noise variables. Hence,
\begin{equation*}
    \text{SNR}_{CS} = \frac{\alpha(u^2+\sigma^2)}{(1-\alpha)\sigma^2}
\end{equation*}
which is independent of $t$. In contrast, for ASCS, the following theorem shows  that $\text{SNR}_{ASCS}^{(t)}$ grows with $t$ until it plateaus.

\begin{theorem}
\label{theorem_3}
Let the sampling threshold designed as $\tau^{(t)} = \tau^{(T_0)} + \frac{\theta}{T}(t-T_0)$ for $T_0 \leq t \leq T$, and set $\tau^{(T_0)} = 0$ and $T_0 = c T$, where $c \in (0,1)$ is a fixed constant. The count sketch only has one hash table. Given any $\theta \in (0, u)$ and $\delta^* > 1-p_0$, then there exists a sufficiently large $T^{'}$, such that for any total sample size $T \geq T^{'}$, the SNR of ASCS is lower bounded by,
\begin{eqnarray}
\text{SNR}^{(t)}_{ASCS} 
\geq 
\frac{1-\delta^*}{\Phi\left(-\frac{\theta(\sqrt{t} -\sqrt{T_0})}{\kappa_0\sigma}\right) p_0 + 1-p_0} \cdot
\text{SNR}_{CS},
\label{eq_4}
\end{eqnarray}
where $\kappa_0$ and $p_0$ is defined in theorem~\ref{theorem_1}.
\end{theorem}

\textbf{Interpretation:}
Theorem~\ref{theorem_3} shows that when $T$ is large enough, as the sampling procedure continues, the ratio $(\text{SNR}^{(t)}_{ASCS}/\text{SNR}_{CS})$ increases almost exponentially fast to its limit $(1-\delta^*)/(1-p_0)$. Note that $1-p_0$ encodes the probability of colliding with signal variables, which is close to 0 when signals are sparse. $\delta^*$ is the probability of missing a signal variable during the sampling, which is controlled by Algorithm~\ref{alg:algo3}. Thus, $((1-\delta^*)/(1-p_0)) \gg 1$ when $T$ is sufficiently large. Therefore, after we start sampling, ASCS achieves a much larger $\text{SNR}^{(t)}_{ASCS}$ compared to vanilla CS. 
Similar to the previous theorems, theorem~\ref{theorem_3} requires ASCS to only have one hash table. For ASCS with \textbf{multiple hash tables}, the LHS of equation~\ref{eq_4} is lower bounded. The lower bound can be approximated similar to the RHS of equation~\ref{eq_4}, with $\kappa_0$ replaced by $\kappa = \sqrt{1 + \frac{\pi(p-1)(1-\alpha)}{2K(R-\alpha)}}$, and $p_0$ replaced by $p^K_0$.

\vspace{-0.2cm}
\subsection{Relaxation of Problem Statement}

Our Problem statement and designed theory assumes (1) the same expectation for all the signals and (2) the same variance for all $X_i$s. Both conditions are rarely strictly satisfied in practical covariance matrix sketching problems. 
We can relax those assumptions as follows:
\begin{enumerate}
    \item Instead of assuming the same expectation for all the signal variables, we can use a lower bound of the signal strength to replace $u$. If we can lower bound the expectation of the signal strength by $u$, theorem~\ref{theorem_1} and~\ref{theorem_2} continue to ensure that ASCS does not miss too many signals during the sampling.
    \item Instead of assuming the same variance for all the $X_i$s, we can replace $\sigma^2$ with the average of the $\Var{X_i}$. Empirically, we can explore the first $r$ samples and calculate the average of their $L_2$ norm square, i.e. approximate $\EX{\Var{X_i}}$ by mean of $\Var{X_i} \approx \frac{1}{pr} \sum_{t=1}^{r}\sum_{i=1}^{p} X^{(t)2}_i$.
\end{enumerate}
In the following subsection, we will implement both relaxations of model assumptions to evaluate the theorems.

\vspace{-0.2cm}
\subsection{Validation of theorems}

Theorem~\ref{theorem_1} and ~\ref{theorem_2} upper bound the proportion of missing a signal covariance during the sampling procedure. Theorem~\ref{theorem_3} lower bounds the SNR ratio of ASCS over CS, $(SNR^{(t)}_{ASCS}/SNR_{CS})$. We also test whether the theorems hold in the real experiments. Again, we use the same method in section~\ref{sec_4.5} to generate a simulation dataset and bootstrap the ``gisette'' dataset($2,000$ samples and $1,000$ features). The size of hash table  is set to $R=p/20$ and the number of hash tables $K=5$. Our formal theorems only considers $K=1$, but using multiple hash tables is more common, which is also the setup of section~\ref{sec: experiments}. In the previous sections, we have extended the theorems to multiple hash tables case while the probability bounds are approximated with close form expressions (rigorous probability bound do not have close form solutions). In this section, we will use those approximated close form bounds for evaluation.

Theorem~\ref{theorem_1} upper bounds the probability of missing a signal covariance at time $T_0$ (when we start to sample the covariances). We set the upper probability of missing a signal covariance (denoted as $\delta$ in section~\ref{sec4.1}) with a range of values from  $0.05$ to $0.1$. Table~\ref{table_1} suggests that observed real probability of missing a signal covariance at time $T_0$ is strictly smaller than the corresponding bound $\delta$.  Theorem~\ref{theorem_2} upper bounds the probability of missing a signal covariance between $T_0$ and $T$. Similarly, we vary the upper bound of this probability ($\delta^* - \delta$). Again, table~\ref{table_1} suggests the real probability is upper bounded by theorem~\ref{theorem_2} at different levels. 

\vspace{-0.4cm}
\begin{table}[ht]
\resizebox{\linewidth}{!}{
\begin{tabular}{|c|c|c|c|c|c|c|}
\hline
target $\delta$            & 0.05   & 0.06   & 0.07   & 0.08   & 0.09   & 0.10   \\ \hline
real prob           & 0.0056 & 0.0089 & 0.0097 & 0.0117 & 0.0126 & 0.0174 \\ \hline
target $\delta^* - \delta$ & 0.05   & 0.07   & 0.09   & 0.11   & 0.13   & 0.15   \\ \hline
real prob           & 0.0421 & 0.0538 & 0.0551 & 0.0595 & 0.0696 & 0.0717 \\ \hline
\end{tabular}
}
\resizebox{\linewidth}{!}{
\begin{tabular}{|c|c|c|c|c|c|c|}
\hline
target $\delta$            & 0.05   & 0.06   & 0.07   & 0.08   & 0.09   & 0.10   \\ \hline
real prob                  & 0.0140 & 0.0184 & 0.0188 & 0.0220 & 0.0240 & 0.0341 \\ \hline
target $\delta^* - \delta$ & 0.10   & 0.12   & 0.14   & 0.16   & 0.18   & 0.20   \\ \hline
real prob                  & 0.0204 & 0.0212 & 0.0268 & 0.0280 & 0.0397 & 0.0421 \\ \hline
\end{tabular}
}
\caption{Comparison between the target probability bound $\delta$ and $\delta^* - \delta$ calculated from our theorems, and the corresponding real probabilities observed on the simulation dataset (Top) and ``gisette'' dataset (Bottom). The real probability is well bounded by its target probability bound. }
\label{table_1}
\vspace{-0.2cm}
\end{table}

\vspace{-0.3cm}
\begin{figure}[ht]
\centering 
\begin{subfigure}{0.49\linewidth}
  \includegraphics[width=\textwidth]{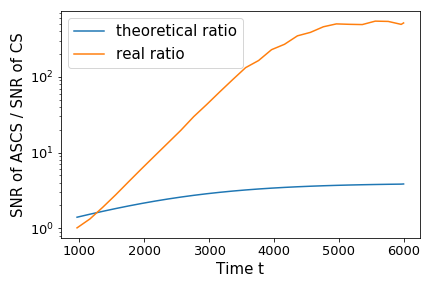}
\end{subfigure}
\begin{subfigure}{0.49\linewidth}
  \includegraphics[width=\textwidth]{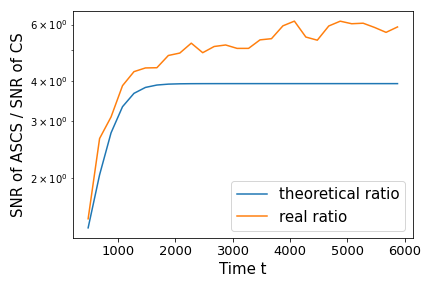}
\end{subfigure}
\caption{The SNR ratio of ASCS over CS that is computed from theorem 3 and tested in real experiments. (Left) simulation dataset (Right) "gisette" dataset. The real SNR ratio is upper bounded by the theoretical SNR ratio.}
\label{fig:SNR_ratio}
\vspace{-0.5cm}
\end{figure}

Theorem~\ref{theorem_3} lower bounds the ratio of $SNR^{(t)}_{ASCS}/SNR_{CS}$ (ROSNR) when $t$ is sufficiently large. Here, to ensure $t$ is large enough, we set the number of samples $T=6,000$ (use all the samples of ``gisette'' dataset).
We compute the theoretical ROSNR using theorem~\ref{theorem_3}, and also computed the real ROSNR from the experiments. To set the hyper-parameters of ASCS, we use $\delta = 0.05$ and $\delta^* = 0.15$. The ROSNR is evaluated every 200 samples. Figure~\ref{fig:SNR_ratio} shows that when $t$ is large enough, the theoretical ROSNR increases to a plateau. Moreover, the real ROSNR is well lower bounded by the corresponding theoretical ratio with the gap also increasing as $t$ becomes larger. We also note that the real ROSNR in the simulation dataset is much larger. This is because the signal and noise variables are separated by a larger margin in the simulation dataset.

\vspace{-0.1cm}
\section{Experiments}\label{sec: experiments}

We perform two sets of evaluations of ASCS. First, we evaluate ASCS on two large scale datasets with hundreds of trillion sized covariance matrices to show the scalability of ASCS (\ref{sec:mainexp}). Then, we also choose five relatively smaller datasets to rigorously evaluate various aspects of ASCS (\ref{sec:smallexp}).

\begin{table*}[ht]
\begin{tabular}{|c|c|c|c|c|c|l|l|l|l|}
\hline
\multicolumn{1}{|l|}{dataset} &
  \multicolumn{1}{l|}{dim} &
  \multicolumn{1}{l|}{average $n_z$} &
  \multicolumn{1}{l|}{\# of corr entries} &
  \multicolumn{1}{l|}{samples} &
  \multicolumn{1}{l|}{size of corr matrix} &
  $K$, $R$ setting &
  Memory &
  CS &
  ASCS \\ \hline
\multirow{3}{*}{URL} &
  \multirow{3}{*}{$10^6$} &
  \multirow{3}{*}{120} &
  \multirow{3}{*}{$10^{12}$} &
  \multirow{3}{*}{$10^6$} &
  \multirow{3}{*}{\textbf{20 TB}} &
  $K=5, R=10^6$ &
  \textbf{20MB} &
  0.439 &
  \textbf{0.979} \\ \cline{7-10} 
 &  &  &  &  &  & $K=5, R=5\times10^6$ & 100MB        & 0.980 & 0.987          \\ \cline{7-10} 
 &  &  &  &  &  & $K=5, R=10^7$   & 200MB        & 0.992 & 0.989          \\ \hline
\multirow{3}{*}{DNA} &
  \multirow{3}{*}{$10^7$} &
  \multirow{3}{*}{378} &
  \multirow{3}{*}{$10^{14}$} &
  \multirow{3}{*}{$10^7$} &
  \multirow{3}{*}{\textbf{0.5 PB}} &
  $K=5, R=10^7$ &
  200MB &
  0.023 &
  0.087 \\ \cline{7-10} 
 &  &  &  &  &  & $K=5, R=10^8$   & \textbf{2GB} & 0.347 & \textbf{0.998} \\ \cline{7-10} 
 &  &  &  &  &  & $K=5, R=10^9$   & 20GB         & 0.999 & 0.999          \\ \hline
\end{tabular}
\caption{Mean of top 1000 correlations reported by ASCS and CS on large scale datasets. Average $n_z$ is the average number of non-zero features per sample.}
\label{tab:top1000}
\vspace{-0.4cm}
\end{table*}

\subsection{Hyperparameters and implementation details} \label{para:hyperparameters}

In our implementation of ASCS, the intricate adjustment mentioned in section~\ref{sec:naivecs} is ignored in our implementation.
In the following section, we introduce how to determine the hyperparameters of ASCS.
\begin{itemize}[leftmargin=*]
    \item {\bf Sparsity $\alpha$ and signal strength $u$:} $\alpha$ is the proportion of signal covariances, and $u$ is the signal strength (or lower bound of signal strength). By definition, $u$ is $(1-\alpha)$ percentile of the covariances vector $\vmu$. The choice of $\alpha$ is subjective, but we can choose a reasonable $\alpha$ based on the distribution of $\vmu$ such that magnitude of the top $\alpha$ percentile of $\vmu$ (signals) is much larger than others (noises). 
    
    Since $\vmu$ is not available in practice, we can spend some samples to explore the distribution of $\vmu$. Specifically, we insert the data into vanilla CS and obtain an approximated $\vmu$ (say $\hat{\vmu}$) (in practice, we only need to store the percentiles of $\hat{\vmu}$). Theoretically, $u$ should be the $(1-\alpha)$ percentile of $\vmu$. But practically, since $\vmu$ is unknown, we can choose $u$ as the $(1-\alpha)$ percentile of $\hat{\vmu}$.

    \item {\bf Probability of missing a signal variable $\delta$ and $\delta^*$:} 
    By algorithm~\ref{alg:algo3} and theorem~\ref{theorem_1}, $\delta$ cannot be smaller than the saturation probability (SP). Thus, we set $\delta = \max(1.01 \cdot \text{SP}, 0.05)$ in the experiment (see sec~\ref{sec4.1} for definition of SP); We also set $\delta^*=\delta+0.15$ such that the probability of missing a signal variable after $T_0$ is upper bounded by 0.15.

    \item {\bf Initial sampling threshold $\tau^{(T_0)}$}: Generally, a small initial threshold is advisable to have smaller exploration period. For correlation matrix, we set $\tau^{(T_0)} = 10^{-4}$.
    For covariance matrix, since the scale of covariance signal of a dataset is unknown, we use small percentile (eg. 10\%ile) of $\hat{\vmu}$ to determine $\tau^{(T_0)}$.
    
    \item {\bf Count Sketch parameters $K$ and $R$:} The design of sketch sizes depends on the memory budget. Given a budget $M$ (store $M$ float numbers), we set $K = 5$ and choose range $R = M / K$. In the experiments below (table~\ref{tab:htsetting}), we show that ASCS performance is robust to value of $K$ in reasonable range ($4{-}10$) 
\end{itemize}

\vspace{-0.2cm}
\subsection{Evaluation of ASCS on `Hundreds of Trillions' scale matrices} \label{sec:mainexp}

A lot of current datasets are huge both in dimensions and number of samples. We evaluate the performance of ASCS on correlation matrix estimation using two trillion scale datasets: URL~\cite{chang2011libsvm} and DNA~\cite{vervier2016large} (Table \ref{tab:top1000}). Consider the DNA $k$-mer dataset (generated using $c=1,k=12,L=200,\text{seed}=42$).
One batch of data is of size 64GB (13M samples, 17M features), and storing the whole covariance matrix (144 trillions of unique entries) would cost \textbf{0.5PB} (petabytes) space.
Since the whole empirical covariance matrix is too large to compute and store, we skip the computation $F_1$ scores of both algorithms. Instead, we evaluate the magnitude of the top $1,000$ largest correlation pairs reported by CS and ASCS (mean of the top $1,000$ reported correlation pairs).  

Table~\ref{tab:top1000} shows that to locate a group of large correlation pairs (with correlations close to $1$) of the DNA $k$-mer dataset, vanilla CS costs 20GB  memory space to find $1,000$ correlation pairs with mean correlation $=0.999$. While ASCS only needs \textbf{2GB} space to achieve a similar performance. We also observe similar results on the URL dataset, where ASCS costs only costs 20MB to find $1,000$ correlations with mean correlation $=0.979$. But CS requires 5 times more memory space to achieve similar performance.
Obviously, ASCS significantly outperforms vanilla CS, and scales seamlessly to large datasets.

\textbf{Running Time:} As discussed in section \ref{sec:ascs}, the time complexity of ASCS and CS  to process the data is $\mathcal{O}(T(n_z + n_u)^2)$, where $n_z$ is the number of non-zero features per sample, $n_u$ is the number of features whose mean is not close to zero, and $T$ is the number of samples. Table \ref{tab:top1000} shows $n_z$ of both DNA $k$-mer and URL dataset. In these datasets, the value of $n_u$ is negligible for most variables. 

We implement both algorithms in python on a single V-100 (32GB) GPU. ASCS and CS easily scale to million-dimensional data (trillion-sized covariance matrix). Compared to CS, ASCS only adds a sampling operation, the computation complexity of ASCS is almost the same as CS since the main computation cost comes from insertion and query operations (Algorithm~\ref{alg:algo2} line 6, 10 and 12). The computation time spent on the sampling procedure is almost negligible. For ASCS, it only costs \textbf{18-25 minutes} to run the experiment on the URL dataset (1.5GB, covariance matrix size 20TB). We can process a larger dataset like DNA $k$-mer in about \textbf{12-15 hours}. Our implementation is far from optimized, and the time measurements are not done by allocating exclusive resources. Therefore, though the current running time is already impressive, we believe it still has a huge scope for further improvement.
\vspace{-0.2cm}
\begin{table}[ht]
\begin{tabular}{|c|c|c|c|}
\hline
Datasets & data dimension & \# of samples & choice of $\alpha$ \\ \hline
gisette  & 5,000     & 6,000   & 2\%   \\ \hline
epsilon  & 2,000     & 400,000 & 10\%  \\ \hline
cifar10  & 3,072     & 50,000  & 10\%  \\ \hline
sector   & 55,197    & 6,412    & 0.5\% \\ \hline
rcv1     & 47,236     & 20,242  & 0.5\% \\ \hline
\end{tabular}
\caption{datasets used for rigorous evaluation}
\label{tab:ds}
\vspace{-0.5cm}
\end{table}

\subsection{Evaluation of ASCS} \label{sec:smallexp}

In this section, we evaluate ASCS on relatively small datasets for which we can compute the exact covariance matrix from the dataset. Access to exact covariance matrix enables us to perform evaluation on all metrics mentioned in section \ref{sec:probstatement}

\textbf{Experimental Setup :}
We evaluate the performance of ASCS and its sensitivity to hyperparameters on small scale datasets. We choose five  datasets from ``LIBSVM'' (listed in Table~\ref{tab:ds})~\cite{chang2011libsvm}. For rigorous evaluation, we have to restrict the size of exact correlation matrix to be able to compute and store it.
we randomly select 1000 features for the experiments. The number of unique entries in this covariance matrix is $\sim 500K$. 

We choose $\delta, \delta^* \text{ and } \tau^{(T_0)}$ as mentioned above in section \ref{para:hyperparameters}. We choose $\alpha$ for each dataset differently and the choice is mentioned in the Table \ref{tab:ds}. Due to the inherent subjectivity in the choice of $\alpha$ and approximation error introduced by using $\hat{\vmu}$ over $\vmu$, we evaluate ASCS by choosing different values of $\alpha$ and $u$ around originally chosen values to show its robustness. We estimate $\hat{\vmu}$ using the first 5\% of the data. We use $R=20,000$ and $K=5$ in this subsection (i.e. memory of sketch $= 20\%$ of \# unique entries of $\Sigma$), unless otherwise stated.

\textbf{Evaluation:} \label{para:evaluation} We evaluate ASCS on 1) mean correlation of the top correlation pairs reported by the sketch; 2) accuracy of locating signals using $F_1$ score. Different users of ASCS might be interested in different top fractions of the signal covariances. To avoid fixing any specific set of signals for evaluation, we just evaluate for varying top fractions of signals.

\begin{figure*}
\centering 
\begin{subfigure}{0.33\textwidth}
  \includegraphics[width=\linewidth]{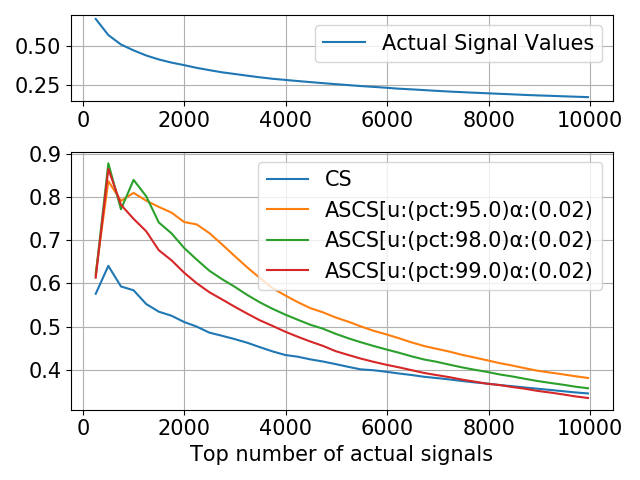}
  \caption{gisette}
  \label{fig:1}
\end{subfigure}\hfil 
\begin{subfigure}{0.33\textwidth}
  \includegraphics[width=\linewidth]{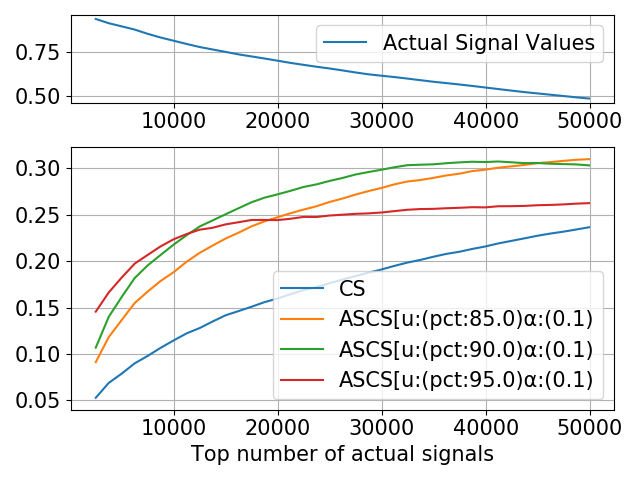}
  \caption{epsilon}
  \label{fig:2}
\end{subfigure}\hfil 
\begin{subfigure}{0.33\textwidth}
  \includegraphics[width=\linewidth]{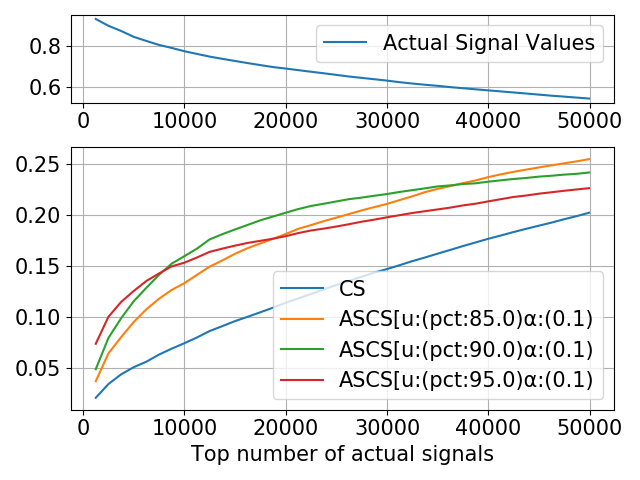}
  \caption{cifar10}
  \label{fig:3}
\end{subfigure}

\begin{subfigure}{0.33\textwidth}
  \includegraphics[width=\linewidth]{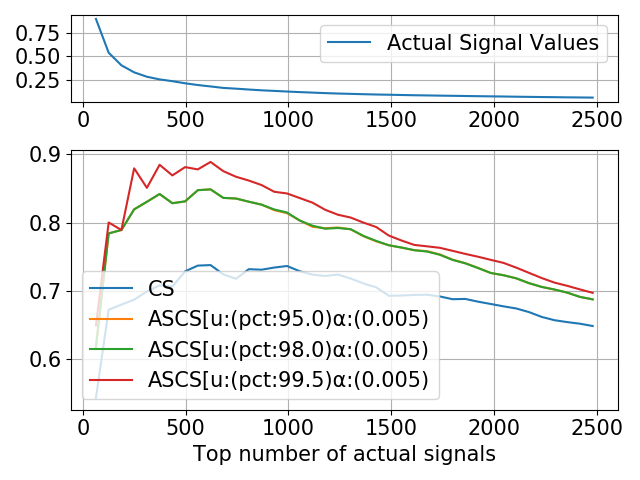}
  \caption{sector}
  \label{fig:4}
\end{subfigure}\hfil 
\begin{subfigure}{0.33\textwidth}
  \includegraphics[width=\linewidth]{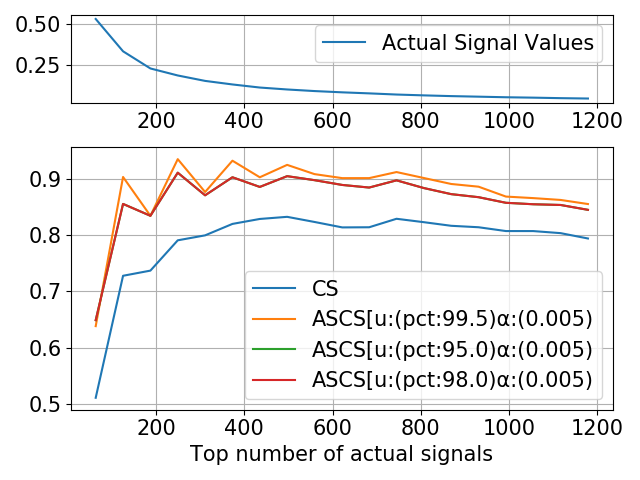}
  \caption{rcv1}
  \label{fig:5}
\end{subfigure}\hfil 
\begin{subfigure}{0.33\textwidth}
  \includegraphics[width=\linewidth]{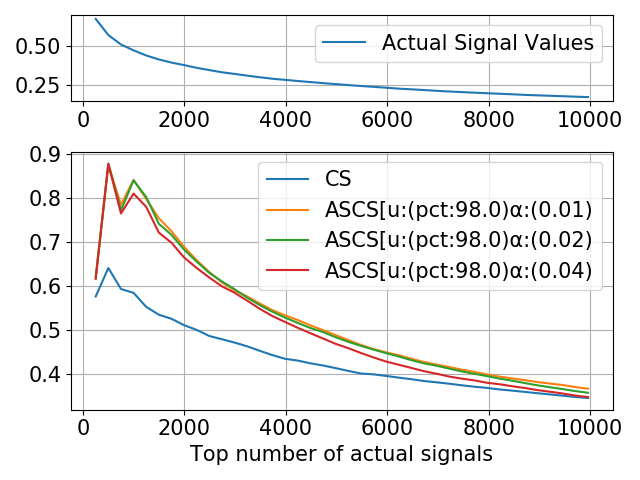}
  \caption{gisette - varying $\alpha$}
  \label{fig:6}
\end{subfigure}
\caption{Accuracy of locating the top signal correlations. Panel (a)-(e) show the robustness of ASCS to choice of $u$. Panel (f) shows the robustness of ASCS to choice of $\alpha$ on ``gisette'' dataset. Labels on x-axis show the number of the top signal correlations and the corresponding correlation values (in brackets). The y-axis corresponds to the maximum $F_1$ score achieved by ASCS and vanilla CS.}
\label{fig:f1max}

\end{figure*}

In Table~\ref{tab:corr}, ASCS shows significant improvement over CS and Augmented Sketch (ASketch \cite{roy2016augmented}) in identifying sets of correlation entries with large mean correlation. Also, Figure~\ref{fig:f1max} suggests that the accuracy of ASCS in identifying signals for correlation is superior to CS as measured by $F_1$ score.

\vspace{-0.3cm}
\begin{table}[ht]
\begin{tabular}{|c|c|c|c|c|c|c|}
\hline
\multicolumn{1}{|l|}{Fraction}   & Algo    & cifar10       & epsilon       & gisette       & rcv1          & sector        \\ \hline
\multirow{3}{*}{$0.01\alpha p$}  & CS      & 0.43          & 0.43          & 0.92          & 0.85          & 0.90          \\ \cline{2-7} 
                                 & ASketch & 0.40          & 0.38          & \textbf{0.98} & 0.85          & 0.88          \\ \cline{2-7} 
                                 & ASCS    & \textbf{0.58} & \textbf{0.62} & 0.97          & \textbf{0.97} & \textbf{0.94} \\ \hline
\multirow{3}{*}{$0.05 \alpha p$} & CS      & 0.39          & 0.39          & 0.62          & 0.54          & 0.71          \\ \cline{2-7} 
                                 & ASketch & 0.36          & 0.36          & \textbf{0.72} & 0.54          & 0.71          \\ \cline{2-7} 
                                 & ASCS    & \textbf{0.53} & \textbf{0.58} & 0.70          & \textbf{0.60} & \textbf{0.78} \\ \hline
\multirow{3}{*}{$0.1\alpha p$}   & CS      & 0.37          & 0.37          & 0.50          & 0.41          & 0.56          \\ \cline{2-7} 
                                 & ASketch & 0.36          & 0.34          & \textbf{0.63} & 0.41          & 0.56          \\ \cline{2-7} 
                                 & ASCS    & \textbf{0.51} & \textbf{0.54} & 0.59          & \textbf{0.44} & \textbf{0.62} \\ \hline
\multirow{3}{*}{$0.25\alpha p$}  & CS      & 0.35          & 0.32          & 0.32          & 0.23          & 0.35          \\ \cline{2-7} 
                                 & ASketch & 0.34          & 0.31          & \textbf{0.47} & 0.23          & 0.35          \\ \cline{2-7} 
                                 & ASCS    & \textbf{0.47} & \textbf{0.47} & 0.36          & \textbf{0.24} & \textbf{0.37} \\ \hline
\multirow{3}{*}{$0.5\alpha p$}   & CS      & 0.33          & 0.29          & 0.22          & 0.14          & 0.23          \\ \cline{2-7} 
                                 & ASketch & 0.33          & 0.29          & \textbf{0.34} & \textbf{0.15} & 0.23          \\ \cline{2-7} 
                                 & ASCS    & \textbf{0.43} & \textbf{0.41} & 0.24          & 0.14          & \textbf{0.24} \\ \hline
\multirow{3}{*}{$\alpha p$}      & CS      & 0.32          & 0.26          & 0.16          & 0.08          & 0.14          \\ \cline{2-7} 
                                 & ASketch & 0.32          & 0.27          & \textbf{0.23} & 0.09          & 0.15          \\ \cline{2-7} 
                                 & ASCS    & \textbf{0.37} & \textbf{0.32} & 0.16          & 0.09          & 0.15          \\ \hline
\end{tabular}
\caption{Mean correlation of top fraction of "$\alpha p$" entries where $p$ is the total number of unique correlation entries and $\alpha$ is the sparsity (proportion of signal correlations, see section \ref{para:hyperparameters})}
\label{tab:corr}
\vspace{-0.3cm}
\end{table}

\textbf{ASCS is robust to the choice of $u$:} 
In Figure~\ref{fig:f1max}(a)-(e), we plot multiple curves for ASCS with different values of signal strength $u$ (eg. $90\%ile$ , $95\%ile$, etc) around $(1-\alpha)$ percentile of $\hat{\vmu}$. We see ASCS performs better than vanilla CS for a wide range of $u$. This shows that improvement offered by ASCS is robust to the choice of different signal strength $u$

\textbf{ASCS is robust to the choice of $\alpha$:} In Figure~\ref{fig:f1max}(f), we plot multiple curves for ASCS with different choices of $\alpha$ and keeping $u$ fixed at original value. We show for ``gisette'' dataset that the $F_1$ score achieved by ASCS is robust to choice of $\alpha$. We restrict to one dataset for lack of space and would add more plots for other datasets in the full version of the paper. 

 \textbf{ASCS vs CS at different sketch sizes:}
 We evaluate ASCS using different sketch sizes (varying $R$) on ``gisette'' dataset. We observe that ASCS consistently outperforms CS across different values of $R$ ($=1,000$ to $100,000$) with K=5. When $R$ ($=100,000$) is large, the improvement is minuscule as CS itself suffers from less collision error and has a very good $F_1$ score. Also, at very small $R$ ($=1,000$) size, hash tables are too crowded and both CS and ASCS have bad $F_1$ scores. For reasonable values of $R$ (eg. $10,000$ or $20,000$), the improvement is significant. (The figures are excluded for lack of space)

\textbf{Sensitivity to the Count Sketch parameters $(K,R)$:}
ASCS requires to set parameters of count sketch : range  $R$ and the number of the repetitions $K$. Given the memory budget allocated to the algorithm, i.e. the total memory budgets is $M$ floating numbers, we may choose to find a suitable $K$, and then the hash table size $R$ can be determined correspondingly by $R = M/K$. We did experiments to evaluate the effect of different K on the performance ASCS for the "gisette" dataset (Table \ref{tab:htsetting} shows the mean correlation of top $0.1\alpha p = 0.2\%$ signals found by ASCS). 

\begin{table}[ht]
\begin{tabular}{|c|c|c|c|c|c|}
\hline
budget $M$ & $K=2$ & $K=4$ & $K=6$ & $K=8$ & $K=10$ \\ \hline
$10K$        & 0.10  & 0.14  & 0.12  & 0.13  & 0.14   \\ \hline
$20K$        & 0.19  & 0.22  & 0.26  & 0.25  & 0.25   \\ \hline
$50K$        & 0.32  & 0.41  & 0.46  & 0.49  & 0.50   \\ \hline
$100K$       & 0.43  & 0.60  & 0.62  & 0.61  & 0.61   \\ \hline
$500K$       & 0.54  & 0.63  & 0.63  & 0.63  & 0.63   \\ \hline
\end{tabular}
\caption{Mean correlation of $0.1 \alpha p$ 0.2\% signals selected by ASCS for ``gisette'' dataset under different hash table settings}
\label{tab:htsetting}
\vspace{-0.4cm}
\end{table}

Table~\ref{tab:htsetting} suggests that ASCS is robust to the choice of $K$ when $K$ is between 4-10. So, if $K$ is not too small or too large, ASCS is not significantly affected by the choice of $K$. Hence, in our experiments, we choose $K=5$, which performs well in different datasets. We want to stress that we did not try to tune ASCS but want to show comparative results between ASCS and CS under reasonable settings.

\textbf{Comparison to other baselines}
ASCS improves the count sketch and incorporates a sampling process to raise the signal-to-noise ratio. Other methods such as Augmented Sketch (ASketch) \cite{roy2016augmented} and Cold filter \cite{zhou2018cold} also try to improve the accuracy of count sketch for frequent elements.  We compared ASCS to Augmented Sketch on the five datasets shown in Table \ref{tab:ds} (We skip the comparison between ASCS and Cold Filter due to its similarity to Augmented Sketch). Table~\ref{tab:corr} shows that compared to Augmented Sketch, ASCS has a better or comparable accuracy of identifying the large correlation entries.
\vspace{-0.2cm}
\begin{table}[ht]
\begin{tabular}{|c|c|c|c|c|c|}
\hline
Dataset & gisette & rcv1 & sector & cifar10 & epsilon \\ \hline
CS      & 47s     & 16s  & 5s     & 41s     & 24s     \\ \hline
ASCS    & 44s     & 13s  & 4s     & 47s     & 30s     \\ \hline
\end{tabular}
\caption{Time comparison in seconds for runs of ASCS and CS on different datasets}
\label{tab:time_comparison}
\vspace{-0.4cm}
\end{table}

\textbf{Execution Speed of ASCS and other baselines}
All the algorithms including ASCS and other baselines such as augmented sketch, naive count sketch are streaming algorithms and have similar execution speeds. Minor differences in computation on each sample of the stream have negligible effect on execution speed. A comparison of time in seconds for ASCS and CS for sketching the data is provided in the Table~\ref{tab:time_comparison}. Timing experiments were done with 29 CPU cores exclusively allotted to each run.


\section{Conclusion}

We propose ASCS for online sparse estimation of a large covariance matrix. Our experiments show that ASCS significantly reduces the memory requirements, and can accurately locate large correlation pairs from matrices with trillions of entries. We also provide theoretical guarantees for ASCS. We envision that ASCS will be widely implemented in different fields and motivate further research on online compression of large scale matrices.

\section{Proof Sketch of theorems}

\textbf{Proof Sketch of Theorem~\ref{theorem_1}:} 
We ignore the sign hash function used in ASCS (assume $s(i)=1$) because the distribution of a noise variable is symmetric around $0$. Adding a sign hash function does not change the distribution. For a signal variable $X_i$, since we consider $\abs{\hat{\mu}^{(T_0)}_i}$, it does not matter whether we use sign hash value $s(i) = +1$ or $-1$. Hence, to simplify the proof, we just ignore the $s(i)$ and it does not affect the result of theorem~\ref{theorem_1}.

Let event $A$ denote the event $\left(\hat{\mu}^{(T_0)}_i < \tau^{(T_0)} \mid \mu_i = u\right)$ which means the estimate of a signal covariance entry falls below threshold $\tau^{(T_0)}$ at time $T_0$.
Let event $B$ denote the signal covariance variable $X_i$ does not collide with another signal covariance variable in the same bucket (for all $j: h(j)=h(i)$, we have $\mu_j = 0$). 

Obviously, the LHS of eq~\ref{eq1} $\leq \PR{A}$, and $\PR{A}$ can be expressed as $\PR{A} = \PR{A\mid B}\PR{B} + \PR{A\mid B^c}\PR{B^c} \leq \PR{A\mid B}\PR{B} + \PR{B^c}$. First, we evaluate $\PR{B}$. For a random covariance variable $X_j$ ($j \neq i$), the probability that $X_j$ belongs to a signal covariance variable and $X_j$ collides with $X_i$ is
$\PR{h(i)=h(j), \mu_j = u} = \PR{h(i)=h(j)}\PR{\mu_j = u} = \frac{\alpha}{R}$, where $h(.)$ is the hash function. So, $\PR{\cup_{j \neq i} (h(i) \neq h(j), \text{or } \mu_j \neq u)} = \prod_{j \neq i} \PR{h(i) \neq h(j), \text{or } \mu_j \neq u} = (1-\frac{\alpha}{R})^{p-1} = p_0$ 

In the next step, we compute $\PR{A\mid B}$. We can express
$\hat{\mu}^{(T_0)}_i 
= 
\frac{T_0}{T}(\bar{X}^{(T_0)}_i + H^{(T_0)}(i))$ and $H^{(T_0)}(i) = \sum_{j:h(j) =h(i)} \bar{X}^{(T_0)}_j \cdot I(i \neq j)$.
Conditional on event $B$, the tail probability of $H^{(T_0)}(i)$ is upper bounded by
$
\PR{H^{(T_0)}(i) \geq \eta \mid B}
\leq 
\Phi\left(-\frac{\eta}{\sigma}\sqrt{\frac{T_0(R-\alpha)}{(p-1)(1-\alpha)}}\right)$, where the probability bound is the tail probability of a normal variable. 
Moreover, since $(\bar{X}^{(T_0)}_i \mid \mu_i=u) \sim N(u, \sigma^2/T_0)$, the tail probability of $\bar{X}^{(T_0)}_i + H^{(T_0)}(i)$ can also be upper bounded by that of a normal variable.
\begin{align*}
\PR{A \mid B}
=&
\PR{\hat{\mu}^{(T_0)}_i \leq \tau^{(T_0)} \mid \mu_i = u, B}  \\
=&
\PR{\bar{X}^{(T_0)}_i + H^{(T_0)}(i) \leq \frac{T\tau^{(T_0)}}{T_0} \mid \mu_i = u, B} \\
\leq& \Phi\left(-\frac{u-\frac{T\tau^{(T_0)}}{T_0}}{\sigma \sqrt{\frac{1}{T_0} + \frac{(p-1)(1-\alpha)}{T_0(R-\alpha)}}}\right).
\end{align*}

Combining with the fact that $\PR{B} = p_0$, $\PR{A} \leq \PR{A|B}p_0 + (1-p_0) \leq \Phi\left(-\frac{u-\frac{T\tau^{(T_0)}}{T_0}}{\sigma \sqrt{\frac{1}{T_0} + \frac{(p-1)(1-\alpha)}{T_0(R-\alpha)}}}\right) p_0 + (1-p_0)$. $\square$

\textbf{Proof Sketch of Theorem~\ref{theorem_2}:} 
Similar to the proof of theorem~\ref{theorem_1}, here, we also ignore the sign hash function of ASCS algorithm due to the same reasons.

Let event A denote the event that \\
$\left(\exists t \leq (T_0, T], \hat{\mu}^{(t)}_i \leq \tau^{(t)}, \hat{\mu}^{(T_0)}_i > \tau^{(T_0)} \mid \mu_i=u, I(i)=0\right)$. Since the LHS of equation~\ref{eq2} $\leq \PR{A}$, we consider upper bound the LHS of equation~\ref{eq2} by upper bounding $\PR{A}$. 
Let $P_x$ be the function of \\
$\PR{\exists t \leq (T_0, T], \hat{\mu}^{(t)}_i \leq \tau^{(t)} \mid \hat{\mu}^{(T_0)}_i = x, \mu_i=u, I(i)=0}$, 
$f_{\hat{\mu}^{(T_0)}_i}(x)$ be the probability density of  $\left(\hat{\mu}^{(T_0)}_i = x \mid \mu_i=u, I(i)=0 \right)$. 
Obviously, LHS of equation~\ref{eq2} $\leq \PR{A} = \int_{x>\tau^{(T_0)}} P_x f_{\hat{\mu}^{(T_0)}_i}(x) dx$. 

To upper bound $\PR{A} = \int_{x>\tau^{(T_0)}} P_x f_{\hat{\mu}^{(T_0)}_i}(x) dx$, we need to bound $P_x$. 
Let $S_i = \{X_j: j \neq i, h(j) = h(i)\}$ be variables colliding with $X_i$. $\hat{\mu}^{(t)}_i = \frac{t}{T}(\bar{X}^{(t)}_i + H^{(t)}(i))$ and $H^{(t)}(i) = \sum_{j:h(j) =h(i)} \bar{X}^{(t)}_j \cdot I(i \neq j)$. 

From time $T_0$ to $t$, it is hard to know how many variables in $S_i$ is not sampled and when they are not sampled. No matter whether $X_j \in S_i$ is sampled or not, $\EX{H^{(t)}(i)} = 0$ always holds true since $\EX{X_j} = 0$. However, $\Var{H^{(t)}(i)}$ will decrease if more $X_j \in S_i$ are not sampled (from time $T_0$ to $t$). Moreover, to upper bound $P_x$, a large $\Var{H^{(t)}(i)}$ leads to a large probability bound of $P_x$. Hence, we may just consider the worst case, where the variables in $S_i$ are always sampled until time $t$. In this case, $\Var{H^{(t)}(i)}$ is larger than any other scenario, and thus, the tail probability of $H^{(t)}(i)$ and $\hat{\mu}^{(t)}_i$ are upper bounded by, $\PR{H^{(t)}(i) \geq \eta \mid I(i) = 0} \leq \Phi\left(-\frac{\eta}{\sigma}\sqrt{\frac{t(R-\alpha)}{(p-1)(1-\alpha)}}\right)$ and 
$\PR{\hat{\mu}^{(t)}_i \leq \tau \mid I(i)=0} \leq \Phi\left(-\frac{u-\frac{T\tau}{t}}{\sigma \sqrt{\frac{1}{t} + \frac{(p-1)(1-\alpha)}{t(R-\alpha)}}}\right)$.

In the next step, we invoke following lemma~\ref{lemma_1} to upper bound $P_x$.
1) Since $\hat{\mu}^{(t)}_i - \tau^{(t)}$ is a submartingale, given $\hat{\mu}^{(T_0)}_i$, $(\hat{\mu}^{(T_0)}_i - \tau^{(T_0)}) - (\hat{\mu}^{(t)}_i - \tau^{(t)})$ is also a supermartingale for $t > T_0$. Hence, $\exp\left(\hat{\mu}^{(T_0)}_i - \tau^{(T_0)} - (\hat{\mu}^{(t)}_i - \tau^{(t)})\right)$ is a non-negative supermartingale. 
2) Since we have bounded tail probability of $\hat{\mu}^{(t)}_i$ by that of a normal variable, in a further step, by lemma~\ref{lemma_1}, we can bound $P_x$. Let $S_t = (\hat{\mu}^{(T_0)}_i - \tau^{(T_0)}) - (\hat{\mu}^{(t)}_i - \tau^{(t)})$.
\begin{align}
P_x 
=& 
\PR{\min_{t>T_0} \hat{\mu}^{(t)}_i \leq \tau^{(t)} \mid \hat{\mu}^{(T_0)}_i = x,  \mu_i = u, I(i)=0} \nonumber \\
=&
\PR{\max_{t>T_0} S_t \geq x - \tau^{(T_0)} \mid \hat{\mu}^{(T_0)}_i = x, \mu_i = u, I(i)=0} \nonumber  \\
\leq&
\frac{\max_{t>T_0} \Exp{S_t \mid \hat{\mu}^{(T_0)}_i = x, \mu_i = u, I(i)=0}}{\Exp{x - \tau^{(T_0)}}} \nonumber \\
\leq&
\Exp{-\frac{2(x-\tau^{(T_0)})(u-\theta)T}{\sigma^2 \left(1 + \frac{(p-1)(1-\alpha)} {T^2(R-\alpha)}\right)}}  \label{eq_5}.
\end{align}

Then, with equation~\ref{eq_5}, we can upper bound $\PR{A}$,

\begin{align*}
\PR{A}
\leq&
\int_{\tau^{(T_0)}}^{\infty} \Exp{-\frac{2(x-\tau^{(T_0)})(u-\theta)T}{\omega^2}} f_{\hat{\mu}^{(T_0)}_i}(x) dx  \\
=& 
\exp\left[\frac{(u-\theta)(\tau^{(T_0)} - \frac{T_0}{T}\theta)}{\omega^2} \right] 
\Phi\left(\frac{T_0(2\theta-u)-\tau^{(T_0)}T} {\sqrt{T_0} \omega}\right).
\end{align*}

Moreover, since LHS of equation~\ref{eq2} $\leq \PR{A}$, LHS of equation~\ref{eq2} is upper bounded by theorem~\ref{theorem_2}. $\square$

\begin{lemma}
\label{lemma_1}
Let $\{Y_t\}^{N}_{t=1}$ be a non-negative martingale/submartingale/ supermartingale with respect to filtration $\{\mc F_t\}^{N}_{t=1}$. Then, for any $t \leq N$ and $\lambda >0$, we have,
\[
\PR{\max_{t \leq N} Y_t \geq \lambda} 
\leq 
\frac{1}{\lambda} \sup_{s \leq N} \EX{Y_s}
\]
where $s$ is a stopping time.
\end{lemma}

\textbf{Proof Sketch of Theorem~\ref{theorem_3}:}
The proof includes two parts. \\
Part 1: With theorem~\ref{theorem_1} and~\ref{theorem_2}, 
we can prove for any $\theta \in (0, u)$ and $\delta^* > 1-p_0$, the probability of missing a signal variable is always controlled below $\delta^*$ when $T$ is sufficiently large. By theorem 2, when $\tau^{(T_0)} = 0$ and $T_0 = cT$, \begin{eqnarray}
&&\PR{\exists t \leq (T_0, T], \abs{\hat{\mu}^{(t)}_i} \leq \tau^{(t)}, \hat{\mu}^{(T_0)}_i > \tau^{(T_0)} \mid \mu_i=u, I(i)=0}     \nonumber  \\
&\leq&
\exp\left[-\frac{(u-\theta) c\theta}{\omega^2} \right] 
\Phi\left(\frac{\sqrt{cT}(2\theta-u)} {\omega}\right)p_0 + (1-p_0)    
\label{eq_7}
\end{eqnarray}
Hence, there exists a $T^{'}$ and $T^{'}_0 = cT^{'}$ such that \\
$\PR{\exists t \leq (T^{'}_0, T^{'}], \abs{\hat{\mu}^{(t)}_i} \leq \tau^{(t)}, \hat{\mu}^{(T^{'}_0)}_i > \tau^{(T^{'}_0)} \mid \mu_i=u, I(i)=0} \leq \\
\frac{1}{2}(\delta^* - (1 -p_0))$.

Note that the RHS of equation~\ref{eq_7} is monotone decreasing in $T$. Therefore, for any $T \geq T^{'}$, the LHS of equation~\ref{eq_7} is upper bounded by $\delta^*$, thus, part 1 is proved. 

Part 2: prove the $\text{SNR}^{(t)}_{ASCS}$ is lower bounded by the RHS of equation~\ref{eq_4}.
Let $f^{(t)}_S$ and $f^{(t)}_N$ be the fraction of signal and noise variables sampled. Then, 
\begin{align*}
\text{SNR}^{(t)} 
= 
\EX{\norm{\mX^{(t)}_S}^2}/\EX{\norm{\mX^{(t)}_N}^2}
= 
\frac{f^{(t)}_S}{f^{(t)}_N} \cdot
\frac{\alpha (\sigma^2+u^2)}{(1-\alpha)\sigma^2}
\end{align*}
Since from part 1, we know $f^{(t)}_S \geq 1-\delta^*$. Similar to the process of proving theorem~\ref{theorem_1}, we have
\begin{align*}
f^{(t)}_N \leq \Phi\left(-\frac{\theta(\sqrt{t}-\sqrt{T_0})}{\kappa_0\sigma}\right)p_0 + 1-p_0
\end{align*}
Now, we can lower bound $(f^{(t)}_S/f^{(t)}_N)$ by the RHS of equation~\ref{eq_4}.   $\square$

\newpage
\bibliographystyle{ACM-Reference-Format}
\bibliography{reference}
\end{document}